\documentstyle[aps,prd,twocolumn,epsf]{revtex}
\tighten
\begin{document}

% Figure inclusion stuff
\firstfigfalse
\newcommand{\incpsf}[1]{\centerline{\epsfysize=3.0in \epsfbox{#1}}}

\wideabs{
\title{Event Horizons in Numerical Relativity II: Analyzing the Horizon}

\author{Joan Mass\'o${}^{(1,2,3)}$, Edward
Seidel${}^{(1,2,4,5)}$, Wai-Mo Suen${}^{(6,7)}$, Paul Walker${}^{(1,4)}$}

\address{
${}^{(1)}$ Max-Planck-Institut f\"ur Gravitationsphysik,
Albert-Einstein-Institut, Schlaatzweg 1, 14473 Potsdam GERMANY \\
${}^{(2)}$ National Center for Supercomputing Applications,
Beckman Institute, 405 N. Mathews Avenue, Urbana, Illinois, 61801 \\
${}^{(3)}$ Departament de F\'{\i}sica, Universitat de les Illes Balears,
   E-07071 Palma de Mallorca, Spain. \\
${}^{(4)}$ Department of Physics,
University of Illinois, Urbana, Illinois 61801 \\
${}^{(5)}$ Department of Astronomy,
University of Illinois, Urbana, Illinois 61801 \\
${}^{(6)}$ McDonnell Center for the Space Sciences, Department of Physics,
Washington University, St. Louis, Missouri, 63130 \\
${}^{(7)}$ Physics Department, Chinese university of Hong Kong, Hong
Kong \\
}

\date{23 April 1998}
\maketitle
\begin{abstract}
  We present techniques and methods for analyzing the dynamics of
  event horizons in numerically constructed spacetimes.  There are
  three classes of analytical tools we have investigated.  The first
  class consists of proper geometrical measures of the horizon which
  allow us comparison with perturbation theory and powerful global
  theorems.  The second class involves the location and study of
  horizon generators.  The third class includes the induced horizon
  2-metric in the generator comoving coordinates and a set of
  membrane-paradigm like quantities.  Applications to several
  distorted, rotating, and colliding black hole spacetimes are
  provided as examples of these techniques.
\end{abstract}
} % End wideabs

%%%%%%%%%%%%%%%%%%%%%%%%%%%%%%%%%%%
% Introduction
%%%%%%%%%%%%%%%%%%%%%%%%%%%%%%%%%%%
\section{Introduction}
\label{sec:Introduction}

Black holes play an important role in general relativity and 
astrophysics.  They are characterized both by spacetime singularities 
within them and by their horizons that cover the singularities from 
the outside world.  In this paper we develop a set of tools for 
analyzing the dynamics of black hole horizons.

The {\em event horizon} (EH) of a black hole is defined as the
boundary of the causal past of future null infinity $\mathcal{I}^+$.
As such the EH surface is traced out by light rays that never reach
future null infinity and never fall into the black hole singularity.
This surface responds to infalling matter and radiation and to the
gravitational fields of external bodies.  In the membrane paradigm of
black holes, the horizon fully characterizes the dynamical
interactions of a black hole with its surroundings~\cite{Thorne86}.
The important role of the horizon in the study of black holes
motivates us to carry out a systematic study of horizon dynamics in
numerical relativity.

While much work has been done on the properties of stationary black
holes and small perturbations about them, little is known about the
properties of highly {\em dynamical} black hole spacetimes.  For
example, the cosmic censorship conjecture \cite{Penrose73}, which
suggests that spacetime singularities should be clothed by event
horizons, demands study into the existence of horizons. The hoop
conjecture \cite{Thorne72a,Misner73}, which states that a black
hole horizon forms if and only if a matter source becomes
sufficiently compact in all directions, begs the question of how
spherical must a black hole horizon be. Caustics, or singular points
in the congruence of photons tracing out the horizon where new
generators can join the horizon, can occur, but under what conditions
do they appear?  And what are the properties of these caustics? One
would also like to know to what extent one can understand interactions
of black holes with their astrophysical environment in terms of
properties of the EH. Studies of most of these questions have to date only
been made in very idealized circumstances or quasi-stationary
spacetimes.  But aspects of each of these open questions are amenable
to study with the numerical methods we describe.

Due to strong field nonlinearities, black hole horizons are difficult
to study analytically.  Therefore we turn to numerical treatment which
is now routinely able to generate highly dynamical, axisymmetric black
hole spacetimes evolved beyond $t=100M$, where $M$ is the ADM mass of
the spacetime.  Many such axisymmetric studies of highly distorted
rotating and non-rotating black holes and colliding black holes have
been performed in recent
years\cite{Abrahams92a,Shapiro92a,Anninos93b,Brandt94b}.  Three
dimensional black hole evolutions are approaching the accuracy of
axisymmetric calculations\cite{Anninos94c,Daues96a,Camarda97a,Cook97a,Bona98b}.
Together with the ability to find and analyze event horizons, these
simulations provide us with a new opportunity to study black hole
dynamics.

We recently proposed methods for the study of the EH in numerically
generated spacetimes~\cite{Anninos94f}.  In a series of followup
papers, we give details of the methods and their applications
to various black hole spacetimes.  The first paper in this
series\cite{Libson94a}, referred to hereafter as Paper I, detailed the
method for locating the EH in a dynamical spacetime, and showed the
high degree of accuracy with which the EH can be located.  In this
second paper, we focus on the tools constructed for analyzing the
dynamics of the EH.

There are several aims of the present paper.  We show three different
sets of tools that can be used to analyzing the dynamics of the EH,
and how one can construct them in numerical relativity.  We show how
accurately the quantities used in these tools can be constructed with
present numerically generated black hole spacetimes.  We demonstrate
the applicability of these tools to various spacetimes of interest.
In fact, these tools apply immediately to almost all numerically
generated black hole spacetimes we have constructed to date.  This
paper describes the tools that elucidate the physics of the EH, and
the accuracy with which we can (or cannot) evaluate these measures;
the emphasis is not on the physics itself.  The physics we learn using
these tools will be discussed in a later paper in this series.

We have developed and present three sets of tools for analyzing the
EH. First, we present a set of geometric measures of the horizon as a
two dimensional surface in a curved 3D space-like slice of constant
time.  These tools include proper circumferences, proper area,
Gaussian curvature, the embedding of the surface in Euclidean space,
and the embedding history.  Second, we discuss how the horizon
generators can be constructed.  This construction also gives the locus
of generators that will join the horizon in the future at caustic
points on the horizon surface.  Third, we present a set of tools from
the membrane paradigm of black holes\cite{Thorne86} for analyzing the
generators and the physics they contain, such as the horizon 2--metric
in generator co-moving coordinates, $\gamma^H_{ab}$ and quantities derived
from and connected to it, such as the expansion $\Theta^H$, shear
$\sigma^H_{ab}$, surface gravity $g_H$, and Hajicek field
$\Omega^H_a$. 

To illustrate the use of these horizon tools, we apply them to several
spacetimes.  We consider the Schwarzschild and Kerr analytic black
hole spacetimes to show the basic principles involved, and to test the
accuracy of the methods.  Also, we apply them to fully nonlinear,
highly dynamical black hole systems, such as a distorted Schwarzschild
BH, a distorted Kerr BH, and the collision of two black holes (the
Misner data\cite{Misner60}).  Our tools can be applied to almost all
numerical black hole spacetimes we have presently constructed, and should be
applicable to future black hole spacetimes as well.

The structure of this paper is as follows: In Sec.~\ref{sec:location}
we briefly review the method we developed to find the location of the
EH. In Sec.~\ref{sec:surface} we show various ways to extract
important information from the EH surface location in the spacetime,
including studying the topology, area, various circumferences,
Gaussian curvature, and geometric embeddings of the surface.  In
Sec.~\ref{sec:generators} we show how to find the actual generators of
the EH, and the information their paths can bring.  In
Sec.~\ref{sec:membrane} we discuss how one can apply ideas developed
in the membrane paradigm\cite{Thorne86} to
numerically generated black hole spacetimes.  Throughout the paper, we
illustrate these ideas with examples from numerically generated black
hole spacetimes.

%%%%%%%%%%%%%%%%%%%%%%%%%%%%%%%%%%%
% Section 2: Locating
%%%%%%%%%%%%%%%%%%%%%%%%%%%%%%%%%%%

\section{Locating the EH in a Numerically Generated Black Hole Spacetime}
\label{sec:location}
Our method for locating event horizons in numerical relativity was detailed in
Paper I.  In order to define our notation, and because our analysis here
is closely related to our EH finding method, we briefly review it here.
The  essence of the EH finding method can be summarized in four steps:

\noindent ({\em i}) At late times after the dynamical evolution we
seek to analyze (that is, when the black hole spacetime has returned to
approximate stationarity; e.g., after the coalescence of two black holes
or after all incident gravitation radiation has either radiated into
the hole or into the far wave zone) the position of the EH can often
be located approximately. We can identify a region of the late-time
spacetime which contains the EH, which we call the horizon containing
domain (HCD).

\noindent ({\em ii)} We trace the evolution of the HCD backward in 
time by tracing its outer and inner boundaries as null surfaces.  A 
function describing a null surface at $t=t_{f}$, $x^{i}=x_{f}^{i}$

\begin{equation}
f(t=t_{f},x^i_{f})=0,
\end{equation}

\noindent satisfies the equation 
\begin{equation}
\partial_{\mu}f \partial^{\mu}f = 0,
\label{nullsfc}
\end{equation}
 or,
\begin{equation}
\partial_t f = \frac{ - g^{ti} \partial_i f +
\sqrt{(g^{ti}\partial_i f)^2 - g^{tt} g^{ij} \partial_i f \partial_j f}
}{g^{tt}}.
\label{evolve}
\end{equation}
for outgoing surfaces. The fact that this method represents the
position of the EH directly as a function $f(t,x^{i})$ is particularly
convenient in our construction of horizon analysis tools, as we shall
see below.

\noindent({\em iii})The strength of our method stems from the fact
that the inner and outer boundaries of the HCD converge together
quickly when integrated {\em backwards} in time in many cases of
interest.  When the distance between the two boundaries in a time
slice becomes significantly less than the grid separation used in the
construction of the spacetime, we have accurately located the EH.
This condition can often be met through the entire regime of interest
here.

\noindent ({\em iv}) The choice of parameterization of the surface is 
important.  For the axisymmetric spacetimes used as examples in this 
paper, one convenient choice is

\begin{equation}
\label{eqn:ehparam}
f=\eta-s(t,\theta)
\end{equation}

\noindent where $\eta$ is a radial coordinate and $\theta$ is polar 
angular coordinate.  In what follows, we assume that this function 
$f(t,\eta,\theta)$ has been obtained for the numerically constructed 
spacetime. 

In axisymmetric two black hole spacetimes, such as those generated in
\cite{Anninos93b,Anninos94b,Anninos97b}, we can use the
parameterization in Eq.~(\ref{eqn:ehparam}) to trace the event horizon
through the merger phase. In the \v{C}ade\v{z} coordinate system
\cite{Cadez71} where coordinates are centered around each individual
throat and the axis below the throat is a line of constant $\theta$,
this parameterization allows us to trace a single hole by applying an
upwinded condition on the horizon at the axis before coalescence.
In the recently proposed ``Class I'' coordinate
system \cite{Anninos97b} where the coordinates are centered around the
throat and the axis, forming peanut shaped radial coordinate lines near
the throats, this parameterization will represent a null
surface which contains both the horizon and the null surface which
represents the locus of generators waiting to join the horizon; a simple
symmetry boundary condition on the equator suffices. The locus can
also be located in the \v{C}ade\v{z} system by using an alternate $(\rho,z)$
parameterization, as described in Paper I. We will use
simulations generated in both coordinate systems interchangeably here.

%%%%%%%%%%%%%%%%%%%%%%%%%%%%%%%%
% Section 3: Studying
%%%%%%%%%%%%%%%%%%%%%%%%%%%%%%%%
\section{Studying the EH Surface}
\label{sec:surface}
\subsection{Topology of the EH}
We note that one important, but easy to obtain piece of information
contained in $f(t,x^i)=0$ is the topology of the EH at a constant time
slice.  In this section we show an interesting case of the EH
undergoing a change of topology.  We apply our EH finding method above
to a numerically generated spacetime representing the head--on
collision of two equal mass black holes with axisymmetry.  In Fig.
\ref{twobhtopology} we show the function $f(t,x^i)=0$ at two times for
the case of Misner time symmetric initial data, described by two
throats connecting two identical asymptotically flat sheets, evolved
by a code described in Ref.~\cite{Anninos93b} (the ``\v{C}ade\v{z}'' code).
The case considered here is for the Misner parameter $\mu=2.2$, for
which the initial distance between the throats is $8.92M$, where $M$
is the $1/2$ the ADM mass.  For details of the initial data set,
see Ref.~\cite{Misner60,Anninos93b}.

%%-FIG-%% twobhtopology
\vbox{ \begin{figure}  
\incpsf{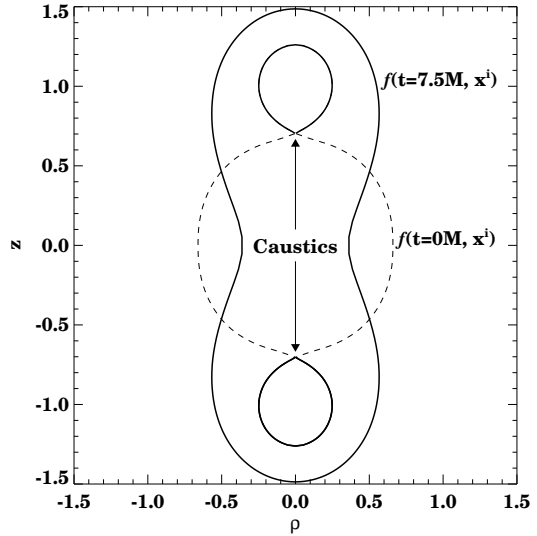}
\caption{We show the topology of the EH for the collision of two black
  holes using data generated with the \v{C}ade\v{z} code from the Misner $\mu
  = 2.2$ initial data.  The solid lines show the EH at $t=0M$
  and $t=7.5M$.  The dotted line at $t=0M$ shows the locus of
  generators which will join the horizon in the future.  We note
  that between $t=0M$ and $t=7.5M$ the horizon undergoes a non-trivial
  topology change. We note that ``crossover'' type caustics form on
  the axis. We shall see below that new generators join the horizon at
  these points.}
\label{twobhtopology}
\end{figure} }

At $t=0$ the EH has the topology of two disconnected two spheres 
represented by the solid lines centered near $z = \pm 1$ in the 
 $\rho-z$ plane.  We note that the function $f(t=0,x^i)$ gives not 
just the location of the EH, but also the locus of the future horizon 
generators before they join the horizon. At $t=7.5M$ the horizon has
the topology of a single sphere.

We treat this change of EH topology by following the surface function
backwards in time.  We can trace the horizon from $t=75$ or $100M$ to
a ``dumbell'' shaped horizon at $t=7.5M$. In Fig.~\ref{twobhtopology},
we start with this ``dumbell'' shaped horizon. Tracing this surface
backward towards $t=0M$, we see that the central part of the surface
shrinks rapidly, and the left and right hand sides cross, indicating
the change in topology.  At $t=0$ the portion of the surface
corresponding to the locus of photons which {\em will} join the
horizon, but have not {\em yet} done so, is given as a dashed line.
As discussed in Sec.\ref{sec:generators}, the crossing of the
surface signals that photons are leaving the horizon, going backwards
in time.  The crossed portion of the surface (shown as a dashed line
in Fig.\ref{twobhtopology}) is no longer on the EH, but represents the
surface of horizon generators ``waiting to be born'', as they will
join the horizon at a future time. For the work here, we define a
point on the horizon where generators cross as a caustic, and
therefore, at the point where the generators cross and join the
horizon, the horizon has a caustic point.  This 
caustic at the cusp in the event horizon is discussed further below, and also
in Ref.~\cite{Matzner95a}.

\subsection{Geometry of the EH Surface} 
\label{geometry} 
The function for the surface $f(t,x^i)$, together with the metric
induced on the surface, gives the intrinsic geometry of the EH,
from which important physical properties can be determined.  In this
section we present a set of tools which allow one to study the
intrinsic properties of the surface.

\subsubsection{Area}
\label{area}
There has been extensive study of the surface area of black hole event
horizons in general relativity
\cite{Hawking73a}.  The
area plays a central role in the thermodynamics of black holes.  As
area is a quantity directly used in analytic studies, it is important
to be able to study the dynamical evolution of the area of the EH in a
numerically constructed spacetime, both for understanding the
spacetime, and also as a diagnostic tool for the accuracy of the
numerical treatment.

Construction of the surface area as function of time is 
straightforward.  Here we show how one computes the area mainly for 
establishing the notation used in this paper.  A surface 
$f(t_0,x^i)=0$ determines the coordinate location 
$x^i=x^i(\bar{\theta},\bar{\phi}), (i=1,2,3)$, of the surface at time 
$t_0$, where we regard the surface as being parametrized by two 
surface coordinates $\bar{x}^a = (\bar{\theta},\bar{\phi}), (a=1,2)$.  
Denote the spatial line element of the spacelike hypersurface at 
$t=t_0$ by

\begin{equation}
d\sigma^2 = g_{ij} \,dx^i \,dx^j,
\label{3metric}
\end{equation}
and so we can define an induced horizon 2--metric as

\begin{equation}
\label{eqn:surfmetric}
\gamma_{ab} = g_{ij} \frac{\partial x^i}{\partial \bar{x}^a}\frac{\partial
x^j}{\partial \bar{x}^b}.
\end{equation}

\noindent The surface area at a time $t$ is then given by

\begin{equation}
\label{eqn:sfcarea}
A(t) = \int{}\sqrt{\gamma} \,d\bar{x}^1\,d\bar{x}^2,
\end{equation}

\noindent where $\gamma$ is the determinant of $\gamma_{ab}$.

In Fig.~\ref{schwarzarea} we show $A(t)$ for a Schwarzschild black
hole evolved with maximal slicing.  The dotted line (labeled Standard ADM)
shows the results obtained using the standard numerical treatment as
described in Ref.\cite{Bernstein89}.  In this case, the calculation is
carried out using a 1D code with 200 grid zones.  It is well known
that when evolved with such slicings and without shift, the EH will
expand outward in the radial direction in coordinate space.  At the
same time, a sharp peak in the radial metric function develops near
the EH.  Due to numerical error caused by the inability to resolve
this sharp peak, the function $A(t)$ deviates significantly from the
analytic value of $16\pi M^2$ as the evolution continues.

%%-FIG-%% schwarzarea
\vbox{ \begin{figure}
\incpsf{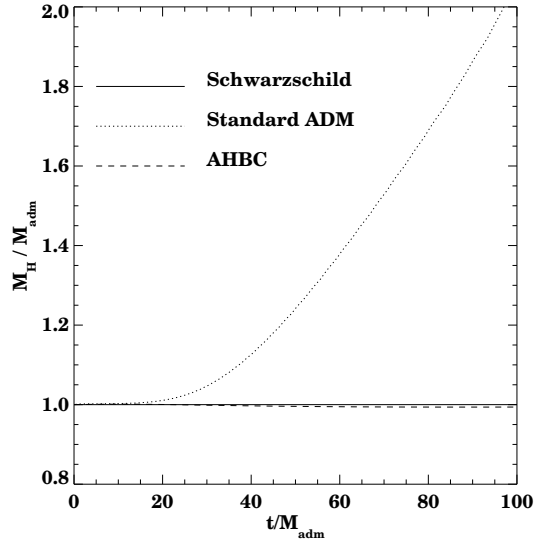}
\caption{We show the area of the EH traced out for a
Schwarzschild spacetime evolved with two different methods.  The dotted
line shows the EH evolution with a standard 3+1 ADM evolution scheme.
The dashed line shows the evolution with an apparent horizon boundary
condition.  The solid line shows the analytic value.  We note that the
same method is used for both spacetimes.  The error in the standard ADM
spacetime is due to inaccuracies in the spacetime metric, not the EH finder.
}
\label{schwarzarea}
\end{figure} }

We compare this result to the case of the dashed line in
Fig.~\ref{schwarzarea}, which is obtained by applying
Eq.~(\ref{eqn:sfcarea}) to a Schwarzschild spacetime constructed with the
same grid parameters, but with an apparent horizon boundary condition
\cite{Seidel92a,Anninos94e}.  The improvement in accuracy is dramatic.

We stress that the issue here is the accuracy of the numerically
constructed spacetime, and not the accuracy of the EH finding method;
an identical finder is used for both the dashed and dotted line. The
error in $A(t)$ in the case of the dotted line is dominated by the
error in the spacetime data.  The relative numerical error in finding
the function $f(t,r,\theta,\phi)=0$ as the position of the EH is
small compared to the errors in the background spacetime.  The
agreement of the dashed line with the analytic value $A(t) = 16\pi
M^2$ suggests that the error in $A(t)$ determined with the apparent
horizon boundary condition spacetime is only about 1\% at
$t=100M$. Though simple, this is an illustrative example of using the
horizon analysis as tool to understand the accuracy of a given
numerical spacetime.

\subsubsection{Circumference}

For black holes with symmetries, the definitions of some
circumferences are geometrically meaningful.  For example, in
axisymmetric spacetimes one can define a polar circumference $C_p$,
and for spacetimes with a reflection symmetry around the equatorial plane,
an equatorial circumference $C_e$.  In the axisymmetric system, with $\partial/\partial\phi$ being the azimuthal Killing
vector, we can take the horizon coordinates $\bar{x}^a$ to be those
tied to the symmetry axis,
\begin{equation}
\bar{x}^a = (\bar{\theta},\bar{\phi})=(\theta,\phi).
\end{equation}
The polar circumference $C_p$, the circumference of a line
with $\phi=\mathrm{const}$, is
\begin{equation}
C_p = \int_{\mathrm{const} \bar\phi} \sqrt{\gamma_{ab} \,d\bar x^{a} 
\,d\bar x^{b}}.
\label{polar circ}
\end{equation}
The equatorial circumference $C_e$, which is the loop around the horizon at
$\theta=\pi/2$, is given by
\begin{equation}
C_e=\int_{\bar\theta = \pi/2} \sqrt{\gamma_{ab} d\bar x^{a} 
d\bar x^{b}}
\label{equatorial circ}
\end{equation}

What is often more interesting is not $C_p$ or $C_e$ by themselves,
but their ratio $C_r=\frac{C_p}{C_e}$.  This defines an effective
shape parameter for axisymmetric surfaces.  Roughly speaking, if
$C_r>1$ or $C_r<1$ the surface is prolate or oblate, respectively.

\paragraph{Shape of the Analytic Kerr Horizon}
In Fig.~\ref{kerrshape} we show the quantity $C_r$ for Kerr black
holes with various rotation parameters $a$.  The numerical simulation
of such spacetimes have been discussed in
Refs.\cite{Brandt94a,Brandt94b,Brandt94c}.  The Kerr spacetimes we
consider here, however, are not evolved, but rather the analytic
(stationary) Kerr solution in the logarithmic radial $\eta,\theta$
coordinates.  The use of analytic data enables us to test directly the
accuracy of our horizon treatment, without being affected by the error
of representing the spacetime on a numerical grid with finite
resolution. The solid line shows the analytic value\cite{Smarr73b},
and the diamonds are data points obtained by applying our methods to
Kerr spacetimes and measuring their circumferences as described above.
The agreement on the plot is excellent.

%%-FIG-%% kerrshape
\vbox{ \begin{figure}
\incpsf{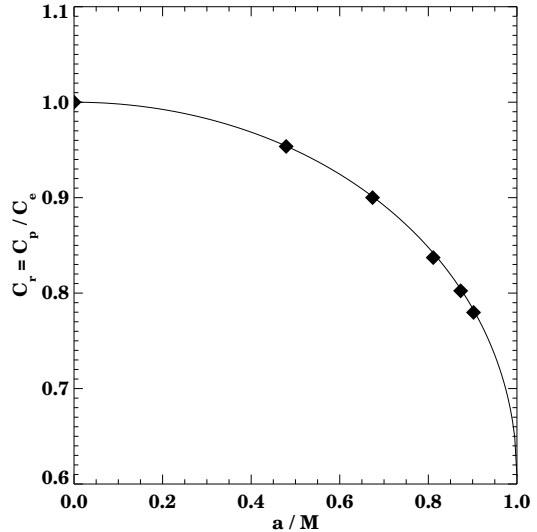}
\caption{We plot the ratio $C_r$ of the polar to equatorial circumferences
of the event horizons for Kerr black holes with various rotation
parameters $a$.  The diamonds show data points obtained by applying our
method to various Kerr spacetimes, while the solid line
shows the analytic result.  The agreement is excellent.}
\label{kerrshape}
\end{figure} }

\paragraph{Shape of distorted EHs}

Next we consider the event horizons of highly distorted black holes.
An important open question about the nature of black holes is the
following: ``Is the event horizon always rather spherical, as
suggested by the Hoop
Conjecture\cite{Thorne72a,Flanagan91a,Flanagan92a}?'' This question
can be addressed to some extent by studying the EH of black holes
distorted by axisymmetric gravitational waves.  The initial data
construction has been described in detail in Ref.~\cite{Bernstein94a}.  For
our purposes it suffices to note that the system corresponds to a time
symmetric torus of gravitational waves, whose amplitude and shape are
specifiable as parameters, that surround an Einstein-Rosen bridge.  In
Fig.~\ref{brillshape} we survey the event horizons at the initial
time $t=0$ for a range of black hole data sets with fixed Brill wave
shape parameters ($(\sigma=1.0, \eta_0=0.0 ,n=2)$ in the language of
Ref.  \cite{Bernstein94a}) representing a quadrupolar wave centered on
the black hole throat with a width of order 1M.  To find the EH at the
initial time, we first evolve the initial data to a late time, and
then trace the EH backwards through the evolved data, as described
above.  Fig.~\ref{brillshape} shows the EH parameter $C_r$ for the
initial data as a function of the Brill wave amplitude $Q_0$.  We see
that in the range of parameters investigated $C_r$ can be rather large
(almost 3 in Fig.~\ref{brillshape}), but does seem to have a maximum
in $Q_0$ space when measured at $t=0$. In contrast, at the same
incident wave amplitude, the AH has much larger amplitude and is
increasing in increasing amplitude to substantially larger distortions
than the EH.  As this paper is restricted to the introduction of the
analysis tools of the EH, we will defer an exhaustive parameter search
and discussion of the physical implications of this result to a later
paper, including a comparison with Fig.4 of \cite{Brandt94a}, where
the upper bound on the distortion of the apparent horizon is found to
be orders of magnitude {\em larger} than that of the EH. In the
initial data, the AH is far inside the EH, but after a short evolution
in these spacetimes, the AH will quickly pop out (the AH is
generically spacelike) to be closer to (but still inside) the EH.

%%-FIG-%% brillshape
\vbox{ \begin{figure}
\incpsf{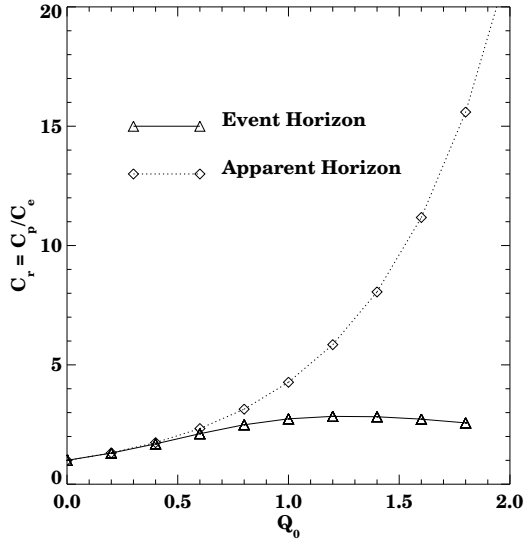}
\caption{We plot the shape parameter $C_r$ for a series of event
  horizons of non-rotating black holes surrounded by gravitational waves
  of varying strengths, denoted as triangles. For comparison, we show
  the behavior of the apparent horizon, denoted as diamonds. We note
  that $C_r$ of the AH at $Q_0 = 1.8$ is around 17, in
  contrast with the EH, which has $C_r$ around 2.5. }
\label{brillshape}
\end{figure} }

While Fig.~\ref{brillshape} shows $C_r$ for highly distorted black
hole event horizons at $t=0$, the same function also provides
important insight into the evolution of these horizons.  In
particular, we study the case of $Q=1.0$. As shown in Fig.~\ref{brilloscillation},
when this black hole evolves, its horizon evolves towards sphericity,
overshoots, and oscillates about its equilibrium, spherical
configuration.  The frequency and decay rate are, to very high
accuracy, the quasinormal mode (QNM) of the black hole as determined by
perturbation theory.  For comparison, a fit to the two lowest QNM
frequencies is given by the dashed line.

%%-FIG-%% brilloscillation
\vbox{ \begin{figure}
\incpsf{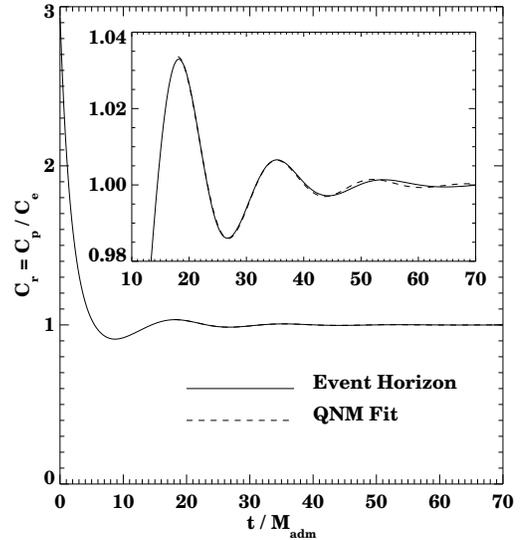}
\caption{The time development of the shape parameter $C_r$ is shown for a
single black hole distorted by a Brill wave.  We note that although
the black hole is initially very distorted ($C_r = 2.9$) it
quickly settles down to the quasi-normal mode ringing  predicted by
perturbation theory.}
\label{brilloscillation}
\end{figure} }

The oscillations of the EH are a common dynamical
feature of black holes.  In Fig.~\ref{2bhoscillation} we show the
oscillation of the EH in the two black hole collision simulated using
the ``Class I'' coordinates with the Misner separation parameter $\mu
= 2.2$,
whose coordinate location was shown in Fig.~\ref{twobhtopology}.  
We show the oscillation of the single horizon which forms after the merger
of the two individual black holes.  The results are similar to
those of the single distorted black hole described above, showing the
generic nature of this phenomenon.

%%-FIG-%% 2bhoscillation
\vbox{ \begin{figure}
\incpsf{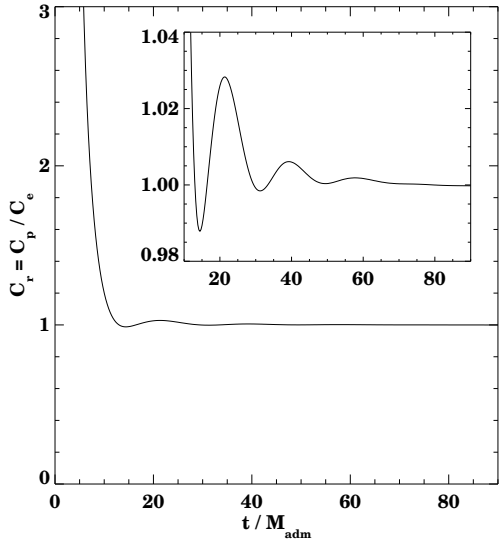}
\caption{The time development of the shape parameter $C_r$ is shown for
  two colliding black holes.  We note again that despite the violent
  initial beginnings,
  the system settles down to ringing behavior at
  late times. Note that the definition of the shape parameter is
  not appropriate until after the coalescence. }
\label{2bhoscillation}
\end{figure} }

\subsubsection{Gaussian Curvature}
\label{Gaussian Curvature}
The Gaussian curvature is a local property describing the two
principal radii of curvature at each point on a surface.  It has been
found to be very useful in analyzing the dynamics of the AH
\cite{Anninos93a}, and it applies equally well to the EH surface.  The
general formula for the Gaussian curvature of a 2-surface in a
spacelike hypersurface with 3-metric $g_{ij}$ is $\kappa = 2 R$ where
$R$ is the Ricci scalar of the 2-sphere with the induced 2-metric.

Fig.~\ref{distorted gauss} shows the time evolution history of the
Gaussian curvature for the highly distorted hole studied in Fig.
\ref{brilloscillation}.  Horizon history diagrams like this have proved very
useful in showing the development of apparent horizon surfaces in
time\cite{Anninos93a}, and here we apply them to the EH surface.  The
figure shows the evolution of the Gaussian curvature as a gray-scale across the
surface as a function of time (horizontal axis). We use the $z-$axis
embedding of the horizon (described below) as the vertical axis in the
plots. $\kappa$ is larger initially near the equator and then
oscillates between the poles and equator.  The checkerboard pattern is
typical of a predominantly $\ell=2$ distortion of the horizon, as
discussed in Ref.~\cite{Anninos93a} (there in the AH case).  The
frequency of oscillation of the horizon surface can be read off
directly from the figure.  We see that it has a period of about $17M$,
which is the fundamental period of the lowest QNM of the black hole.
After about $t=60M$, the hole gradually settles down to its final,
spherical configuration.

%%-FIG-%% distorted gauss
\vbox{ \begin{figure}
\centerline{\epsfxsize=3.0in \epsfbox{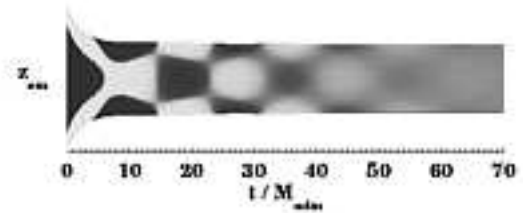}}
\caption{We show the time evolution history of the
  Gaussian curvature for the highly distorted hole ($Q_0 = 1.0$). We
  plot the curvature as a gray scale using the embedded $z$ value as
  the $y-$ axis, and $t/M$ as the $x-$ axis. We note that even though
  the hole is initially very distorted with large curvature, it
  settles down to a damped oscillatory pattern at later times with a
  frequency of about $17M$.  }
\label{distorted gauss}
\end{figure} }

In Figs. \ref{2bhgauss}a and \ref{2bhgauss}b we show a similar diagram
for the two black hole collision (Misner $\mu=2.2$) evolved in ``Class
I'' coordinates.  At late times, $\kappa$ has an appearance similar to
that shown by the highly distorted case discussed above.  At early
times, there are two separate black holes whose horizons are about to
merge.  The surfaces are most highly distorted along the caustic line,
and the Gaussian curvature is largest there.  In Fig.~\ref{2bhgauss}a we
show the entire history of $\kappa$ for this system.  We see that in
the early times, before coalescence, the Gaussian curvature is very
high near the coalescence point.  ($\kappa$ is in fact singular on the
EH at the caustic; we show the curvature very close to the caustic).
In Fig.~\ref{2bhgauss}b we show the early time behavior,
so the details of the curvature can be seen around the
coalescence point.

%%-FIG-%% 2bhgauss
\vbox{ \begin{figure}
\incpsf{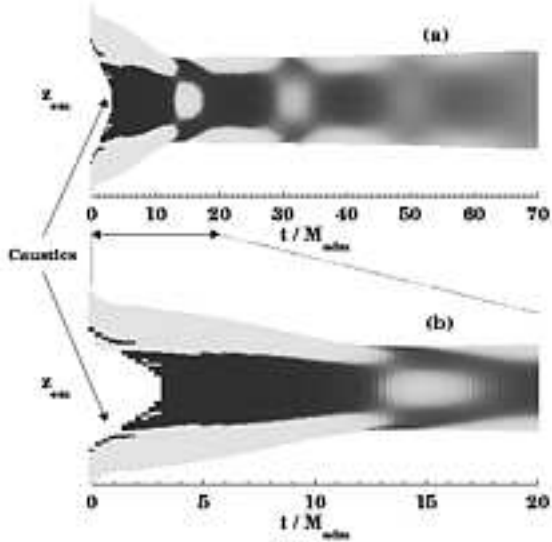}
\caption{We show the time evolution history of the
  Gaussian curvature for the 2BH collision (Misner $\mu=2.2$). In Fig.
  (a) we show the entire history of the horizon curvature, and note
  the repeated oscillation pattern, as in Fig.\protect\ref{distorted
    gauss}, here with a more complicated pattern, but still with a
  frequency of about $17M$. In Fig.~(b) we show the early time
  behavior of the system, seeing the strong curvature near the systems
  cusp as the holes come together. We note that the line of caustics
  is the region inside the two holes before coalescence, as indicated
  by the arrows.}
\label{2bhgauss}
\end{figure} }

\subsubsection{Embedding Diagrams and Embedding Histories}
\label{embeddings}

The use of embedding diagrams to study the intrinsic geometry of
spacetimes is not new in relativity.  It is a particularly useful way
to study a curved 2D surface on a 
constant time slice.  The embedding technique creates a fictitious 2D
surface in a flat 3D Euclidean space with the same geometric
properties as the original 2D surface in the curved 3D space.  This
technique has been described fully in Ref.  \cite{Anninos93a}, where
it was used to study AH surfaces.  We follow the same embedding
approach here.

We can perform a non-trivial test of our embedding treatment by embedding the
analytic Kerr horizon, and comparing this horizon with the embedding
diagrams predicted in Ref.~\cite{Smarr73b}. For high-rotation Kerr black
holes ($a/M > \sqrt 3 / 2$), it is not possible to
embed the horizon around the pole. In Fig.~\ref{kerrembed} we show how
our EH finder finds and embeds the correct analytic horizon up to the
$\theta$ value where the embedding no longer exists.

%%-FIG-%% kerrembed
\vbox{ \begin{figure}
\incpsf{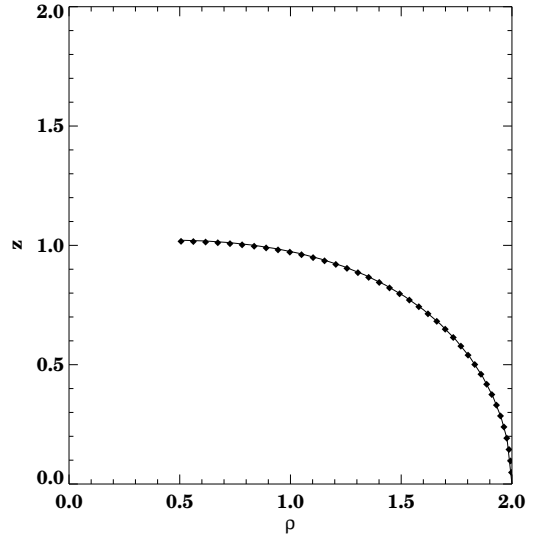}
\caption{We compare the embedding of the Kerr horizon from our horizon
  finder with the known embedding for the value $a/M=0.877$, at which
  the entire horizon cannot be embedded into flat space. We notice the
  agreement between our finder, the diamonds, and the
  known solution, the line.}
\label{kerrembed}
\end{figure} }

In Fig.~\ref{distorted embedding}a-\ref{distorted embedding}c we show
a time sequence of the embedding diagram for the black hole studied in
Fig.~\ref{brilloscillation}.  In Fig.~\ref{distorted embedding}d we
show the EH and Schwarzschild embeddings, where both embeddings are
normalized by the horizon mass. This normalization removes the
spurious area growth caused by errors in the numerical spacetime, as
seen in Fig.~\ref{schwarzarea}, from our horizon embeddings.  In
Fig.~\ref{distorted embedding}a we see that the initial embedding is
very prolate, in concordance with the large value of $C_r$ shown in
Fig.~\ref{brilloscillation} at $t=0$.  We also note that the final
state is indeed a Schwarzschild-like horizon, namely a spherical black
hole characterized solely by its mass.

%%-FIG-%% distorted embedding
\vbox{ \begin{figure}
\incpsf{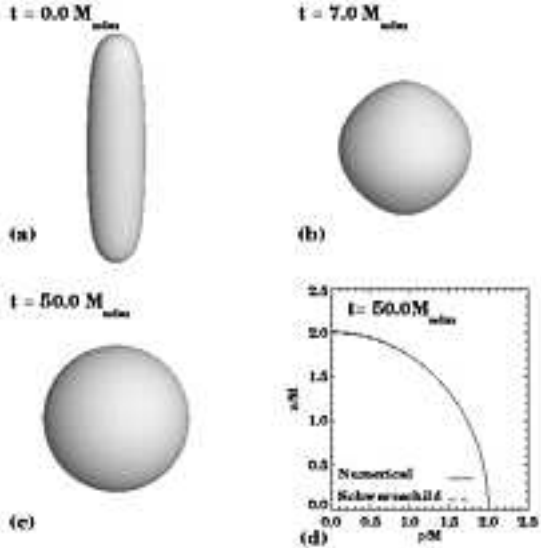}
\caption{We show a sequence of embedding diagrams for an
  EH distorted by a large amplitude Brill wave.  Even though the
  horizon geometry is very non-spherical at $t=0M$, as demonstrated by
  the cigar shaped event horizon, the system quickly becomes fairly
  spherical, as shown in the time snapshots in figures (a), (b), and
  (c), and also in Fig.\protect\ref{brilloscillation}, which
  corresponds to this system. In Fig.~(d) we see that the late time
  horizon is essentially a Schwarzschild horizon, as it has settled
  down to a sphere with radius $2M$.}
\label{distorted embedding}
\end{figure} }

In Figs.  \ref{bowenyork embedding}a-\ref{bowenyork embedding}d we
show a time sequence of embedding diagrams of the EH for a Bowen and
York, rotating black hole \cite{Bowen80,Brandt94a}, with angular
momentum $J=15$, evolved by a code described in Ref.~\cite{Brandt94b}.
We see from Fig.~\ref{bowenyork embedding}a that the initial EH is
quite spherical (We make the front $45^\circ$ of the horizon
transparent to facilitate viewing, hence the ``pacman'' appearance of
the horizon).  The Bowen and York construction differs from that of a
stationary Kerr hole, so this data set can be regarded as containing a
gravitational wave that makes the initial black hole horizon more
spherical than the oblate pure Kerr hole. At $t=12.4M$ into the
evolution, as shown in Fig.~\ref{bowenyork embedding}b, the embedding
has a shape reminiscent of a napkin holder; the top and bottom
sections of the horizon cannot be embedded as they have negative
curvature at the axisymmetric pole, and therefore cannot be represented
in a Euclidean space.  At this instant in time, the extent of the
unembeddable region is near a maximum.  Fig.~\ref{bowenyork
  embedding}c shows the geometry at a later time, as the horizon
settles down towards its Kerr form.  Fig.~\ref{bowenyork embedding}d
shows a quadrant of the EH embedding at time $t=45M$.  At this time,
the hole has settled down to the Kerr form, in accordance with the no
hair theorem \cite{Hawking73a}.  The shape of the EH is to high
accuracy the same as that of an analytic Kerr hole of $a/m = 0.877$,
the embedding of which is plotted as a dotted line for comparison with
the numerical result. We note that the value of $a/m = 0.877$ is the
value of the rotation specified in the initial data solve (a $J=15$
Bowen and York hole), and that this result is still observed late in
the evolution. This is physically required, as an axisymmetric system
cannot radiate angular momentum. The fact that our horizon finder
confirms this late time behavior is a strong verification of the
accuracy of our methods. Notice that there is still a region of the
horizon that cannot be embedded, as the horizon for such rapidly
rotating black hole is ``too flat'' for Euclidean space, and the
regime in which the EH cannot be embedded matches the region for a
Kerr EH, as was also shown in Fig.~\ref{kerrembed}.

%%-FIG-%% bowenyork embedding
\vbox{ \begin{figure}
\incpsf{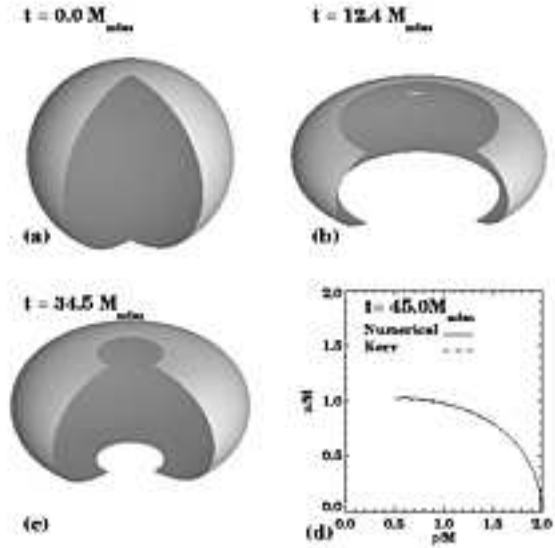}
\caption{We show a sequence of
  embedding diagrams of the EH for a Bowen and York rotating black
  hole, with angular momentum $J=15$ ($a/M = 0.877$). We show
  snapshots of the horizon at (a) time $t=0$ (b) time $t=12.4$, and
  (c) time $t=34.5$. To allow clearer visualization of the region
  where the embedding fails, we make the front $45^\circ$ of the
  horizon transparent, hence the ``pacman'' appearance of the holes.
  We note that, at late times, the system approaches the analytic Kerr
  embedding.}
\label{bowenyork embedding}
\end{figure} }

In Figs.~\ref{misner embedding}a--\ref{misner embedding}d we show four
snapshots of the embedding of the EH for the two black hole head-on
collision case generated with the \v{C}ade\v{z} coordinate system for $\mu =
2.2$.  Fig.~\ref{misner embedding}a shows the embedding of the EH on
the initial, time symmetric slice ($t=0$).  We see the two individual
black holes, with cusps on each horizon on the $z-$axis.  In
Fig.~\ref{misner embedding}b we show the embedding at time $t=5.4$,
shortly after the merging of the two holes.  In Fig.~\ref{misner
  embedding}c we see the late time spherical behavior of the horizon,
despite the system's tumultuous beginnings.  In Fig.~\ref{misner
  embedding}d, we compare the embedding of the EH at $t=80M$, shown as
a solid line, to the horizon of a Schwarzschild hole with the
appropriate mass, shown as a dashed line (once again, we normalize our
final embedding by the final area).  Again we see the no hair
theorem at work, in that the initial condition with no charge and no
angular momentum settled down to a black hole completely described by
its single parameter, $M$.

%%-FIG-%% misner embedding
\vbox{ \begin{figure}
\incpsf{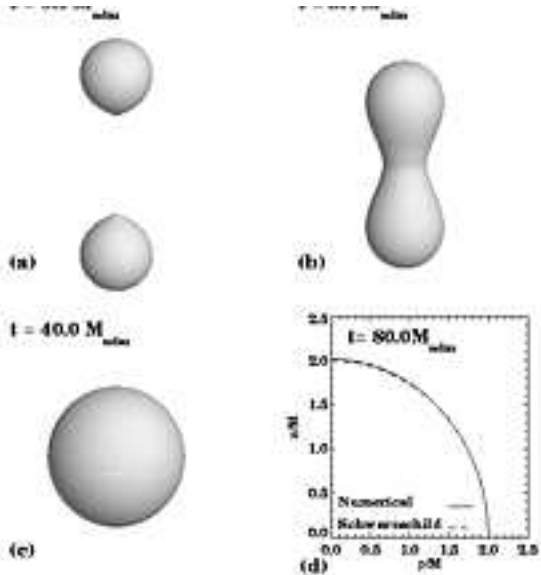}
\caption{We show a sequence of
  embedding diagrams of the EH for the two black hole head-on
  collision case at (a) time $t=0$ (b) time $t=5.4M$, and (c) time $t=40.0M$.
  The simulation here was generated with the \v{C}ade\v{z} code. We note that
  at late times, as shown in (d), the system approaches the appropriate
  Schwarzschild hole, as expected.}
\label{misner embedding}
\end{figure} }

To bring out the dynamics of the horizon evolution, it is useful to
show the ``embedding history diagram'' of the horizon instead of a
series of snapshots.  In Fig.~\ref{2bh embed history}, we show the
evolution of the embedding in time for the two black hole case just
discussed.  In this diagram the $\phi$ direction has been suppressed,
i.e., we stack up $\phi=\mathrm{const}$ cross sections of the 2D
embeddings from various times to create a continuous, 2D embedding
history diagram.  We note that this figure is not a spacetime diagram,
in that the $(\rho,z)$ space away from the horizon surface has no
physical or mathematical connection to the curved $3+1$ spacetime.
However, these embedding history diagrams are a convenient and
effective method for showing the evolution of the embedding of the
event horizon surface in coordinate time ($t$) in the fictitious
Euclidian ($\rho,z$) space.  This figure shows the geometry of the
individual holes as they approach each other, with a cusp on each
horizon.  The distance between the holes before the merger, which is
not prescribed in the embedding process for data generated in the
\v{C}ade\v{z} coordinates, is chosen to keep the embedding history diagram
smooth.  After the merger, one can (barely) see the oscillation of the
final horizon, which occurs at the normal mode frequency of the final
black hole.  In this diagram we also show the evolution of various
horizon generators (photons moving normal and tangent to the horizon)
as lines on the surface.  The determination and use of these
generators will be discussed in detail in the next section.

%%-FIG-%% 2bh embed history
\vbox{ \begin{figure}
\incpsf{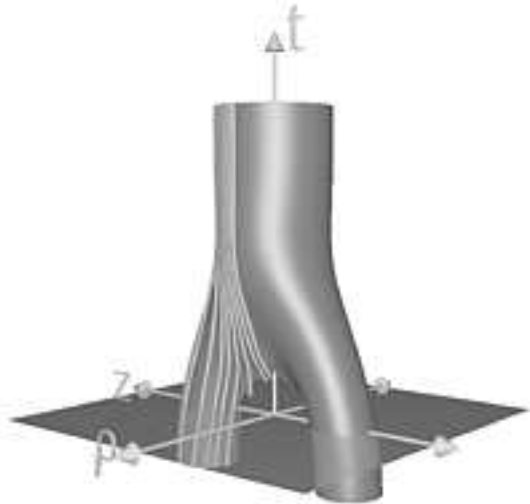}
\caption{We show the embedding history diagram of the two black
  hole collision generated with the \v{C}ade\v{z} code. This diagram, the
  famous ``pair of pants'' diagram, shows a time history of the
  embedding of the horizon by stacking consecutive embeddings on top
  of each other in time. The lines on the surface show the paths of
  the horizon generators, and shows them leaving the surface at the
  crossover caustic, as will be discussed below.}
\label{2bh embed history}
\end{figure} }

Another interesting embedding history diagram is shown in
Fig.~\ref{rot embed history}.  Here we show the embedding of the {\em
  equator} of the horizon of a Bowen and York black hole from
Fig.~\ref{bowenyork embedding}.  We see the equator bulge out and then
back in, as the hole becomes more and less prolate (the total area
increases in time).  As discussed in Sec.~\ref{sec:generators} we
embed the equator since it allows us to show the generator motion in
the $\phi$ direction.

%%-FIG-%% rot embed history
\vbox{ \begin{figure}
\incpsf{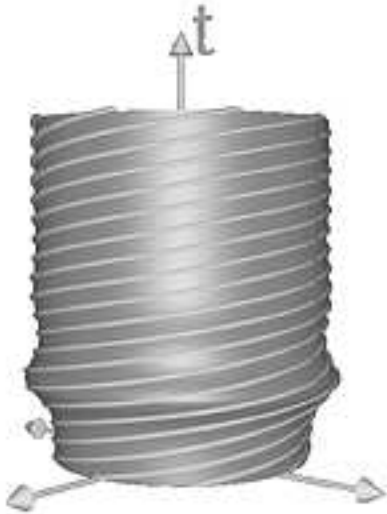}
\caption{We show the embedding history diagram for the case of a
  distorted Kerr black hole. Here we suppress the $\theta$ direction,
  and embed only the equator at all $\phi$ values. Although the
  spacetime is axisymmetric, so there is no $\phi$ variance, this
  embedding demonstrates the rotation of the generators in the $\phi$
  direction as the system evolves in time. This diagram is a numerical
  construction of the ``barber pole twist'' diagram.}
\label{rot embed history}
\end{figure} }

\subsection{Numerical Convergence of the Horizon Measures}

The study of numerical convergence is important for any numerical
treatments based on finite differencing, and we discuss it for each of
our results here. We give a brief overview of numerical convergence
here, but for a more detailed discussion, see
Refs.~\cite{Choptuik91,Bona98a}. To avoid confusion with Paper I, we
emphasize that the numerical convergence we discuss here is the usual
convergence rate of our numerical results depending on grid
resolution. It is a {\em completely} different phenomenon than the
{\em physical} convergence discussed in Sec.~\ref{sec:location} and in
Paper I, which is a physical attraction of null surfaces to the
horizon independent of numerical treatment.

Given three solutions to a discretized equation, $L$, $M$, and $H$ at
resolutions $\Delta x$, $\Delta x / q$, and $\Delta x / q^2$, 
the convergence exponent is defined as 
\begin{equation}
\sigma = \frac{\log \frac{L-M}{M-H}}{\log q}
\label{eqn:sigma_conv_def}
\end{equation}
where $-$ is simple subtraction for numbers, and a combination of
interpolation onto a common grid and reduction via a norm operator for
fields. The measure $\sigma$ indicates that the error in a numerical
solution is of order $\Delta x^\sigma$.

In Fig.~\ref{scalconv} we show the numerical convergence exponents of
the horizon area $A$ and ratio of circumferences $C_r$ in a slightly
distorted single black hole evolving in time ($Q_0 = 0.1$, $\eta_0 =
0$, $\sigma_0 = 1.0$, $n=2$).  We choose this case for our convergence
studies since the spacetime is quite accurate so effects of numerical
error in the background spacetime are minimized, and we can directly
test our horizon treatments (We see similar convergence results for
all the spacetimes discussed here; since we have not assumed the
Einstein Equations hold in any of our analysis so far, constraint
violations in the spacetime will {\em not} affect the convergence of
the system, although they could in principle cause the horizon
analysis to converge to a non-physical result).  The convergence study
is made by keeping the spacetime resolution fixed in all runs and
adjusting only the number of points which represent the horizon.  We
use an interpolator of order equal to or higher than our evolution
method on a numerical grid of data.

We see that the measures $\mathcal A$ and $C_{r}$ converge at second 
order as expected.  These quantities are simply measures 
of the interpolated metric and the surface (evolved with a second 
order MacCormack method) so any result below second order would 
signify an error.

Additionally, we measure (but do not show) the average convergence of
the $z$-coordinate of the embedding over the entire surface (In the
embedding procedure, only the $z$-coordinate is integrated; The
$\rho$-coordinate is exactly given as a function of $z$ and the
metric).  The embedding converges at first order.  This is to be
expected, since we use a first order integration over the derivative
of the surface to form the embeddings.  Since the embedding is only
measured, not evolved, this first order nature is satisfactory. We
note that using a higher order integration scheme would not
substantially improve the accuracy of our embeddings, since we cannot
remove integrals over derivatives of the surface from our embedding
procedure.

%%-FIG-%% scalconv
\vbox{ \begin{figure}
\incpsf{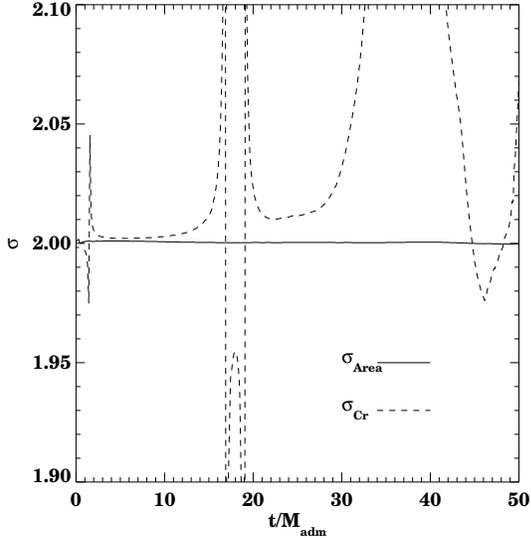}
\caption{We show the time evolution of the convergence exponent of $A$
  and $C_r$ in the low amplitude gravitational wave plus black hole
  spacetime. We note second order convergence throughout the entire
  run, and that the convergence rate of $C_r$ has (small) spikes
  associated with an oscillatory function, while that of $A$ does
  not.}
\label{scalconv}
\end{figure} }

%%%%%%%%%%%%%%%%%%%%%%%%%%%%%%%%
% Section 4: Horizon Generators
%%%%%%%%%%%%%%%%%%%%%%%%%%%%%%%%

\section{Horizon Generators}

We have already seen examples of generators of the horizon in several
of the above figures. In this section we show that the horizon
generators can be located in a numerical spacetime using information
already constructed in our surface-based horizon finder.  We will use
these generators to study the motion and dynamics of black hole event
horizons.

\label{sec:generators}
\subsection{Formulation}
The EH is generated by null geodesics.  With the EH given by $f(t,x^i)=0$,
the null geodesics that generate the surface satisfy
\begin{equation}
\frac{dx^{\alpha}}{dt}=A(x^{\mu}) g^{\alpha \beta} \partial_\beta f,
\label{generator_evolution}
\end{equation}
where $A(x^\mu)$ is a scalar function of the four coordinates.  Notice
that in terms of $f$, the generators satisfy a first order equation,
rather than the more complicated second order geodesic equation.  We
choose the normalization $A(x^\mu)$ to be
\begin{equation}
A(x^{\mu})=\frac{1}{g^{t\beta}\partial_\beta f}
\label{normalization}
\end{equation}
so that the null vector tangent to the null geodesics is given by
\begin{equation}
\ell^\mu = (1,\frac{g^{i\beta}\partial_\beta
f} {g^{t\alpha}\partial_\alpha f}).
\label{eq:gen_ell}
\end{equation}
Notice that with this choice, the null geodesic is {\em not}
affinely parameterized, but instead, adapted to the global time coordinate $t$
used in the numerical calculation of the spacetime itself.

One important advantage of determining the null generator using 
Eqs.~(\ref{generator_evolution}) and (\ref{normalization}) is that in this
formulation, the trajectories obtained are guaranteed to lie on the EH.
This is in contrast to numerically integrating the second order
geodesic equation directly.  As shown in Paper I, integration of the
geodesic equation directly can lead to spurious tangential drifting
effects which can significantly affect the position of the
horizon generators.  This difference can lead to errors of
interpretation, as described in Paper I.  The importance of obtaining
accurate trajectories of the horizon generators is clear.  
Generators of the horizon  contain all the information of the
dynamics of the EH. The entire membrane formulation described
Sec.~\ref{sec:membrane} is based on these trajectories.  Thus,
inaccurate location of the generators due to tangential drifting can
make analysis of the horizon dynamics via the generators impossible.

\subsection{Analytic Kerr as a test case}

We briefly study the motion of generators in the analytic Kerr case.
In this case, a generator will rotate in the $\phi$--direction on the
horizon with a rate
\begin{equation}
\frac{d\phi}{dt} = \frac{a}{2M^2 + 2M(M^2-a^2)^{1/2}}.
\label{analphi}
\end{equation}

In Fig.\ref{analphidot} we show $d\phi/dt$ for various Kerr
spacetimes. We measure the $\phi$ location of the horizon generators,
numerically differentiate in time, and show the result as solid black
diamonds. We compare these results with the analytic result,
Eq.~\ref{analphi}, shown as a solid line. We note the results
agree with the analytically expected value.

%%-FIG-%% analphidot
\vbox{ \begin{figure}
\incpsf{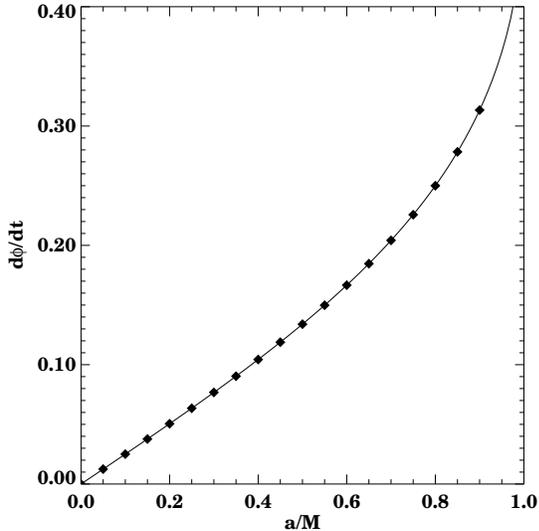}
\caption{We show horizon generator angular velocity, $d\phi/dt$ for a
  $M=1$ black hole with values of $a/M$ between 0 and 1 as solid
  diamonds.  We compare these results with the analytic value, shown
  as a solid line.  We note the excellent agreement}
\label{analphidot}
\end{figure} }

\subsection{Horizon Generators in Dynamical Spacetimes}

We now apply these techniques to the study of the trajectories of the
horizon generators for three numerically constructed dynamical
spacetimes.

We first consider the low amplitude Brill wave plus black hole
spacetime considered above ($Q_0 = 0.1$). In this spacetime we expect
a non-spherical evolution in the generators. Rather than just moving
radially, as the generators would in a dynamically sliced spherical
spacetime, we also expect some non spherical deflection to be
noticeable in the generators. We show this deflection by plotting the
difference between the generator angular location, $\theta_{gen}$ and
the late time generator location, $\theta_0$, versus $\theta_0$
itself, evolving in time. Equatorial plane symmetry requires there is
no deflection at the equator, and axisymmetry require that there is no
deflection at the pole, thus all the generator deflection must occur
between the equator and the pole.  In the intermediate region, the
generators oscillate with a quasi-normal mode frequency with an
amplitude dying down at late times.  In Fig.\ref{smallpert_gen}, we
show the deflection quantity $\theta_{\rm gen} - \theta_0$ evolving in
time, and note that the deflection is small, but displays this
expected behavior.

%%-FIG-%% smallpert_gen
\vbox{ \begin{figure}
\incpsf{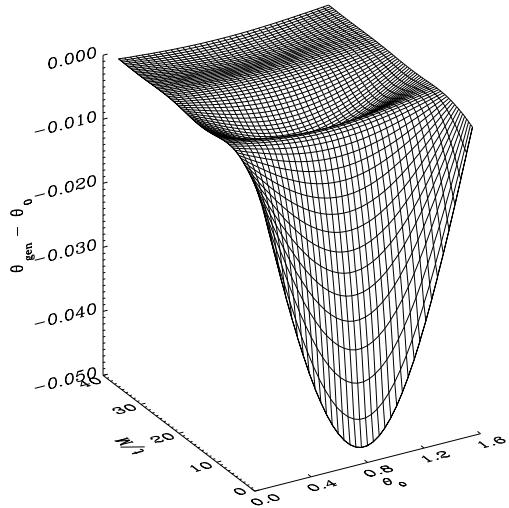}
\caption{We show the angular deflection of generators for a horizon
  with a low amplitude brill wave initially incident on a black hole.
  We show the deflection by plotting the angular location of the
  generators, $\theta_{\rm gen}$ minus their late time position,
  $\theta_0$, versus their late time position, $\theta_0$, evolving in
  time for all generators. From the figure it is clear that the
  angular deflection occurs away from the equator and pole, as is
  obvious from simple symmetry arguments.}
\label{smallpert_gen}
\end{figure} }

In Fig.~\ref{rot embed history}, the ``Barber Pole Twist'' diagram, we
have seen the $\phi$ motion of the generators in Kerr-like spacetimes.
In Fig.\ref{phidot}, we plot the quantity $d\phi/dt$ of the photons
versus  time.  We see that they settle down to a constant value at late
times, with
\begin{equation}
\frac{d\phi}{dt}(t=50M) = 0.293.\nonumber
\end{equation}
This is to be compared to the analytic value of 0.296 given by
Eq.~(\ref{analphi}) with $a/M = 0.877$, denoted by the dashed line in
the figure.  We see that the measured value at $t=50M$ differs from
the analytic value by about $1\%$, which demonstrates that the
hole is settling down to a Kerr black hole at late times, and that the
determination of the horizon generators is quite accurate, although
late time errors in the numerical spacetime lead to the observed small
difference from the expected value in the generator angular velocity.

%%-FIG-%% phidot
\vbox{ \begin{figure}
\incpsf{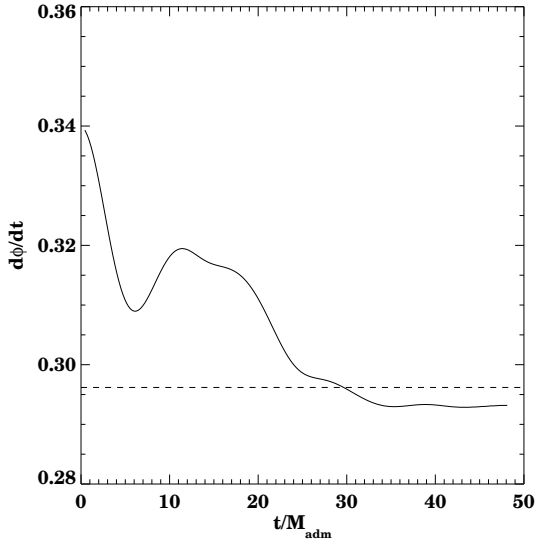}
\caption{We plot the generator angular velocity $d\phi/dt$ for a
  horizon generator on the equator vs. $t$ for the distorted Kerr
  black hole. We note that, although this quantity is not constant in
  time, it approaches the analytic value of $0.296$ at late times, as
  the horizon settles down to its Kerr form.}
\label{phidot}
\end{figure} }

Turning to the ``Pair of Pants'' diagram, Fig.~\ref{2bh embed
  history}, which shows the embedding of two colliding BHs, we see
that the most interesting feature of the generators is that some of
them leave the horizon (going backwards in time) at the inner seam of
the pants.  There is a line of caustic points on the $z$-axis
extending backward from the ``crotch'' point where the two horizons
merge.  It is at these points along the caustic line in the history
diagram that photons originally travelling in the causal past of null
infinity $(J^{-}(\mathcal{I}^{+}))$ join the horizon as generators.
As discussed above, only the surface of the horizon has been embedded;
the photons that have left the embedding diagram have also left the
embedding space, and their paths are only shown to denote their
joining the horizon.

In Fig.~\ref{trajectories} we show the coordinate location of the
generators and horizon surface found using the \v{C}ade\v{z} code.  The EH
location at various times is shown by heavy solid lines. The $t=0$
surface is the horizon of two distinct BHs at the initial time, which
evolves to a single, merged horizon, shown at $t= 3.1M$.  We see that
generators which start outside the EH (denoted by inward pointing
triangles in the figure) move inwards, cross on the $z$-axis, and join
the horizon.  This crossing of generators of the EH in the two black
hole collision is crucial to recent understanding of the structure of
the horizon in the Misner spacetime. Further analysis of the nature of
such lines of caustics is possible and underway\cite{Walker98a}.
Coupled with new techniques for evolving multiple black hole
spacetimes, our techniques should allow an increase in our
understanding of how the generators behave in dynamical multiple black
hole spacetimes.

%%-FIG-%% trajectories
\vbox{ \begin{figure}
\incpsf{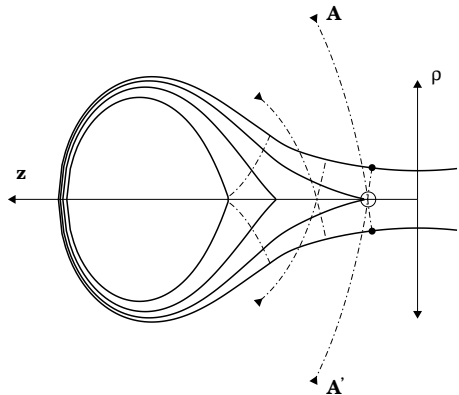}
\caption{In this diagram we analyze the
  trajectories of the photons before and after they join the horizon
  for the case of two colliding black holes (Misner $\mu=2.2$). Slices
  of the horizon are shown at $t=0$, $1.9$, $2.7$, and $3.1M$. We note
  that generators not originally on the horizon (shown by inward
  pointing arrows on this figure) cross over each other at a line of
  caustics on the $z-$ axis and join the horizon as the holes collide.
  For example, the two photons labeled $A$ and $A^\prime$ join the
  horizon at $t = 2.7M$, crossing over at the point shown as an open
  circle. At time $t=3.1M$ they are on the horizon at the points
  shown as solid circles.  The second black hole is not shown, as the
  system has equatorial plane symmetry. }
\label{trajectories}
\end{figure} }

%The Figs.~\ref{2bh embed history} (``Pair of Pants'') and \ref{rot
%embed history} (``Barber Pole Twist'') have been familiar in the minds of
%relativists for the last 25 years.  Now we have succeeded in 
%calculating them for true dynamical black hole spacetimes.  

\subsection{Numerical Convergence of the Generators}

We can measure convergence of the generator locations, just as we
measured convergence of horizon measures. Since the generator location
is an ODE integration with coefficients determined by the surface
location and the derivatives of the surface, the appropriate test is
to keep the number of generators fixed, while changing the spacing of
the surface. We can then measure the differences in generator
locations as a function of spacing of the surface and form a
convergence measure for each generator, which can then be averaged over
all generators.

We show the result of performing this operation on the radial and
angular positions of the generators in Fig.~\ref{genconv}, using the
low amplitude Brill wave plus black hole spacetime considered above.
We note that the radial position of the generators (solid line), which
is non-oscillatory, converges at second order. However, the angular
position (dashed line) has spikes typical of an oscillatory function,
but converges below second order. This lower order convergence is due
to the principal term in the angular generator position evolution
being the (interpolated) derivative of the horizon surface. That is,
since we interpolate second order spatial derivatives of the surface
for the generator sources, the evolution of the generator angular
positions has error terms larger than the $\Delta q^2$ terms. This
convergence order could possibly be increased by using fourth order
spatial derivatives and very high order interpolators.

%%-FIG-%% genconv
\vbox{ \begin{figure}
\incpsf{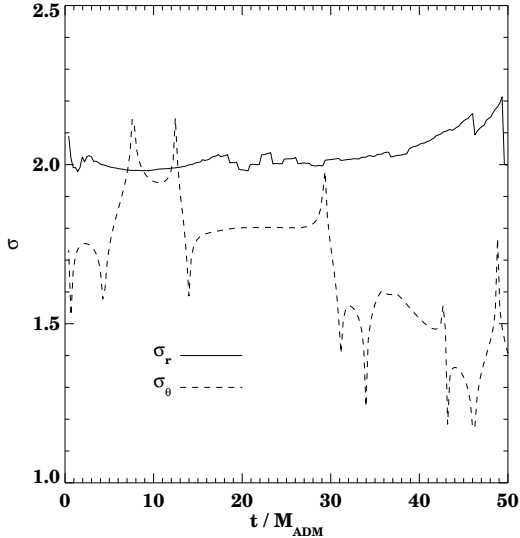}
\caption{We show the convergence of $r$ and $\theta$ for the
  horizon generators in the low amplitude Brill wave spacetime. The
  convergence exponents $\sigma_r$ and $\sigma_\theta$ are plotted
  versus time. We note that the radial location of the generators
  converges at second order, but the angular location converges at
  somewhat less than second order. This is unsurprising since the
  evolution equation for generators' angular location is dominated by an
  interpolation of the numerical derivative of the horizon surface.}
\label{genconv}  
\end{figure} }

\section{The Membrane Paradigm}
\label{sec:membrane}

\def\expn{\Theta}

We now turn to a detailed analysis of the information carried by the
congruence of horizon generators, and the extent to which this can be
used in numerical relativity as a tool to investigate black hole
dynamics. The theoretical basis for this study is based on the
membrane paradigm (MP) \cite{Thorne86}. The MP views the black hole as
a 2-surface in a 3-space with the properties of a viscous fluid. In
many ways, the EH in a dynamical spacetime is like a soap bubble
perturbed by external influences. The MP is particularly valuable in
providing an intuitive understanding of how a BH reacts to its
surroundings.

There has been much study of gravitational interactions using the MP
in quasi-stationary situations~\cite{Suen88,Price86}. With the advent
of numerical identification of the EH and generators described above,
we can now start to consider applying the MP to fully non-linear and
dynamical spacetimes.  With this goal in mind, we demonstrate how to
construct the MP quantities on a numerically located EH, and examine
the accuracy of these constructions in several testbed spacetimes.

\subsection{Formulation}
\label{formulation}
We begin by discussing the MP formalism with the goal of being able to
construct MP quantities on our numerically located horizons.
The membrane paradigm requires the choice of a time slicing,
splitting up spacetime into an ``absolute space'' and a ``universal
time''\cite{Thorne86}.  To apply the MP to numerical relativity, we choose
the universal time to be the same as the time coordinate $t$ used in
the numerical evolution.  This implies that the time coordinate used
in the numerical evolution has to be well behaved on the EH.  This is
the case for all of the black hole spacetimes we have numerically
constructed.

We define the four vector $\vec{\ell}$ to be the tangent to the
horizon generators, and we normalize it as in Eq.~(\ref{eq:gen_ell}) above,
with $t$ being considered as the ``universal time''.  This vector is
in the full 4-dimensional space, which we index with Greek letters,
$\mu, \nu \ldots = (0,1,2,3)$.  On the 2D spacelike section of the EH
at constant $t$, we choose spacelike 2D coordinates $\bar{x^a}$ which
we index with lower case Roman letters, $a,b,\ldots = (2,3)$, which
are {\em comoving} with the horizon generators, i.e., 

\begin{equation}
\vec{\ell} = \frac{\partial}{\partial t} \Bigg\vert_{\bar{x^a}}  =
\frac{\partial}{\partial\bar t}
\label{eq:ell_is_tbar}
\end{equation}
where $\bar t$ is the comoving generator time coordinate (which is
identical to the time in the simulation by Eq.~(\ref{eq:gen_ell})).
In a coordinate basis, we have the spatial basis vectors
\begin{equation}
\vec{e_a} = \frac{\partial}{\partial \bar{x^a}} \Bigg\vert_{t}
\end{equation}
which are orthogonal to $\vec\ell$ by construction. We define the fourth basis vector $\vec{n}$ by
\begin{eqnarray}
\label{eqn:ndef}
\vec{n}\cdot \vec{n}=0 \nonumber\\
\vec{n}\cdot \vec{\ell}=-1\\
\vec{n}\cdot \vec{e}_{a}=0. \nonumber
\end{eqnarray}

The induced metric
on the 2D horizon section is
\begin{equation}
\gamma^H_{ab} = e_a^\mu e_b^\nu g_{\mu \nu}.
\label{generator metric}
\end{equation}

In the membrane paradigm the description of the dynamics of the
horizon is given in terms of the horizon surface gravity $g_H$, the
shear $\sigma^{H}_{ab}$, the expansion $\expn^H$, and the Hajicek
field $\Omega^H_a$.  They are defined as
\begin{eqnarray}
\expn^{H} &=& \frac{1}{2} \frac{\partial}{\partial\bar t}{\mathrm 
ln}\,{\mathrm det}\,\gamma^H
\label{horizon_quant_start}\\
\sigma^H_{ab} &=& \frac{1}{2}\left( \frac{\partial \gamma^H_{ab} }{\partial
\bar t}- \expn^H
\,\gamma^H_{ab} \right)\\
\nabla_{\vec{\ell}}\,\,\vec{\ell} &=& g_H\, \vec{\ell}\\
\label{horizon_quant_gh}
\Omega^H_a &=& -\vec{n}\cdot\nabla_a\vec{\ell}
\label{horizon_quant_end}
\end{eqnarray}

These quantities are dependent on the choice of time coordinate $t$,
as they explicitly involve $\vec{\ell}$ in their definition.  That is,
they are gauge {\em dependent} measures of the horizon dynamics.  In
the formulation of the membrane paradigm given in Ref.
\cite{Thorne86}, a particular time slicing is chosen for a stationary
black hole, e.g., a Kerr black hole.  In this slicing, without
perturbation, $g_H$ and $\Omega_H$ take on special values while
$\expn^H$ and $\sigma^H_{ab}$ vanish.  For small perturbations about a
Kerr horizon, $g_H$ and $\Omega_H$ are first order slicing dependent.
In the formulation given in Ref. \cite{Thorne86}, time slicings of the
perturbed black hole are chosen so that the surface gravity $g_H$
remains unchanged in time.  In our application of the membrane
paradigm to numerical relativity, as we are mostly interested in
highly dynamical and fully nonlinear interactions, we do not put such
restrictions on the time slicing.  Rather, we let the time slicing be
determined by the natural choice of the numerical evolution (maximal
slicing for most cases presented in this paper). We expect that the
new features introduced by different slicings will become familiar
when the formulation is used in more black hole studies, and hopefully
allow further insight into the slicing conditions and numerical
evolutions.

The horizon quantities (\ref{horizon_quant_start}) -
(\ref{horizon_quant_end}) satisfy the following:

\noindent The ``tidal force equation''
\begin{equation}
D_{t} \sigma_{ab}^{H} + (\expn_{H} - g_{H})\sigma_{ab}^{H} = 
-C_{a\mu b\nu}\ell^{\mu}\ell^{\nu} \equiv  -{\mathcal E}^{\mathcal H}_{ab},
\label{eq:tidal_force}
\end{equation}
the ``focusing equation''
\begin{equation}
D_{t} \expn_{H}= g_{H}\expn^{H} - \frac{1}{2}\expn_{H}^{2} -
\sigma^{H}_{ab} \sigma_{H}^{ab} - 8\pi T_{\mu\nu}\ell^{\mu}\ell^{\nu},
\label{eq:focusing}
\end{equation}
and the ``Hajicek equation''
\begin{eqnarray}
D_{t} \Omega_{a}^{H} + (\sigma_{a}^{Hc} + 
\frac{1}{2}\delta_{a}^{c}&\expn_{H}&)\Omega_{c}^{H} + 
\expn_{H}\Omega_{a}^{H} = \nonumber\\
 (g_{h} + \frac{1}{2}&\expn_{H}&)_{,a}
-{\sigma_{a}^{Hb}}_{||b} + 8\pi T_{a\mu}\ell^{\mu}.
\label{eq:hajicek}
\end{eqnarray}
$D_t = \perp \ell \cdot \nabla$ is the projection of the
covariant derivative along $\ell$ into the horizon section.  ``$||$''
denotes covariant differentiation on the horizon section.
$C_{\mu\nu\rho\sigma}$ is the Weyl tensor and $T_{\mu\nu}$ is the
energy-momentum tensor. 

The comparison of Eqs.~(\ref{eq:tidal_force})-(\ref{eq:hajicek}) with
the evolution equations for a 2D viscous fluid gives meaning to the
horizon quantities Eq.~(\ref{horizon_quant_start}) -
(\ref{horizon_quant_end}).  One finds that Eq.
(\ref{eq:tidal_force}) describes the response of a fluid to a
gravitational tidal field, Eq.~(\ref{eq:focusing}) describes the
energy conservation of the viscous flow, and Eq.~(\ref{eq:hajicek}) is the
corresponding Navier-Stokes equation of the fluid flow. The
surface density of the mass-energy of the fluid is identified as
$-\expn_H/8\pi$, the surface pressure is $g_H/8\pi$, and the momentum
density corresponds to $-\Omega^H_a/8\pi$.  The dynamics of the EH of a
black hole can be understood in analogy to the motion of a fluid on a
soap bubble.  In the following section, we show how these ``fluid''
quantities can be constructed for an EH located in a numerical
simulation.

\subsection{Constructing Membrane Quantities}
\label{constructing membrane}
Once $f(t,x^i)=0$ is given, we obtain $\vec{\ell}$ as given in Eq.~(\ref{eq:gen_ell})
in a straightforward manner.  Next, we define the comoving coordinates
$\bar{x}^a, (a=1,2)$ on
the horizon section by $(\bar{\theta},\bar{\phi})$.  Then we have
\begin{equation}
\vec{e}_a = \left(
\partial_{\bar{\theta}},\partial_{\bar{\phi}}\right) =
\left(\vec{p},\vec{q}\right).
\end{equation}
The coordinate components of the two basis vectors can be obtained by
\begin{equation}
\vec{p} = \partial_{\bar{\theta}} = p^r \partial_r + p^{\theta}
\partial_{\theta} + p^{\phi} \partial_{\phi},
\end{equation}
where $p^r$ is defined to be
\begin{eqnarray}
p^r &=& \frac{\partial r}{\partial \bar{\theta}} \nonumber \\
    &=& \frac{\hbox{difference in }r\hbox{ for
neighboring generators}}{\hbox{difference in }\bar{\theta}\hbox{ for
neighboring generators}},
\end{eqnarray}
and likewise for $p^{\theta}$, $p^{\phi}$. We use this definition in a
discrete fashion, differencing over generator locations, and
therefore our basis vectors will always have a discretization
error based on the initial spacing of generators in $\bar\theta$
space.  As the coordinates $\bar{\theta}$ and $\bar{\phi}$ are chosen
to be comoving, we have $p^t=0=q^t$. For the axisymmetric cases
considered here, we pick $\phi = \bar\phi$ and thus $\vec q = \partial
/ \partial \phi$, the azimuthal killing vector.

The horizon two--metric is then written as
\begin{equation}
\gamma^H_{ab} = \left(\matrix{\gamma^H_{\bar{\theta}\bar{\theta}}&
                         \gamma^H_{\bar{\theta}\bar{\phi}}\cr
                         \gamma^H_{\bar{\theta}\bar{\phi}}&
                         \gamma^H_{\bar{\phi}\bar{\phi}}}\right).
\label{horizon_2_metric}
\end{equation}
The individual components are defined by, e.g.,
\begin{equation}
\gamma^H_{\bar{\theta}\bar{\theta}} = g_{ij}p^ip^j
\label{eqn:hzmet_forinstance}
\end{equation}

Solving for $\vec n$ is particularly troublesome. We use the
following, geometrically motivated, method.
When we solved for $\partial_t f$ in Eq.~(\ref{evolve}), we solved the
quadratic equation choosing the positive root for the outgoing null
surface. We could also have chosen the negative root, and found
an evolution equation for the {\em ingoing} null surface. Let us call
the ingoing evolution equation $\partial_t^- f$, and Eq.~(\ref{evolve})
$\partial_t^+ f$ temporarily. We will use the notation
$\partial_\mu^{\pm} f = (\partial_t^{\pm} f, \partial_i f)$. Thus, we
can form two null vectors $\vec L$ and $\vec N$ as
\begin{eqnarray}
L_\mu &=& (\partial_t^+ f, \partial_i f) \nonumber \\
N_\mu &=& (\partial_t^- f, \partial_i f).
\end{eqnarray}
From Eq.~(\ref{nullsfc}) it is clear that both $\vec L$ and $\vec N$
are null, and that $\vec L$ is simply $\vec \ell$ with a different
normalization.

However it is also clear that $\vec N \cdot \vec e_a = 0$. To see
this, recall that $\vec e_a$ has only spatial components, so
\begin{equation}
N_\mu  e_a^\mu = e_a^i N_i = e_a^i \partial_i f = e_a^i L_i =
L_\mu  e_a^\mu = 0
\end{equation}
since $\vec L$ is proportional to $\vec \ell$ which is orthogonal to
$\vec e_a$ by construction.

So now all that remains is to find a normalization such that $\vec n
\cdot \vec \ell = -1$. This is straightforward. Since $\vec \ell =
\vec L / A(x^u)$, using Eq.~(\ref{normalization}) it is clear that
\begin{equation}
\vec N \cdot \vec \ell = \frac{g^{\mu\nu} \partial_\mu^+ f
\partial_\nu^- f}{g^{t\alpha} \partial_\alpha^+f} \equiv B(x^\mu)
\end{equation}
and so we can define $\vec n$ by rescaling $\vec N$ by $B(x^\mu)$,
\begin{equation}
\vec n = -\vec N / B(x^\mu).
\end{equation}
We note we can use Eq.~(\ref{eqn:ndef}) to measure how accurately
$\vec n$ and $\vec \ell$'s orthogonality with $\vec e_a$ is
maintained.

Once the horizon 2--metric $\gamma^H_{ab}$ and full set of comoving
vectors, $(\vec l, \vec n, \vec p, \vec q)$ are obtained, we can form
the expansion, shear, and Hajicek Field via Eq.~(\ref{horizon_quant_start}) -
(\ref{horizon_quant_end}).  From Eq.~(\ref{horizon_quant_gh}), 
the surface gravity is
\begin{equation}
g_H = \Gamma^t_{\mu\nu} \ell^{\mu} \ell^{\nu},
\label{sfcgravdef}
\end{equation}
for our particular parameterization of $\vec\ell$.

There are several terms in the definitions of the membrane quantities
which require careful numerical and analytical treatment in order to
be evaluated in our framework.  In particular, in order to evaluate
the horizon quantities accurately, we must be able to evaluate
$\partial \gamma_{ab} / \partial \bar t$, preferably without taking
numerical time derivatives.  From Eq.~(\ref{eqn:hzmet_forinstance}),
the horizon 2--metric $\gamma$ has two types of terms, those due to the
comoving basis vectors $\vec p$ and $\vec q$, and those due to the
spacetime 4-metric, $g_{ij}$.  Thus using the chain rule to evaluate
$\partial \gamma_{ab} / \partial \bar t$ will yield terms like
$\partial g_{ij} / \partial \bar t$ and $\partial p^i / \partial \bar
t$.

The derivatives along the generators of the spacetime metric can be
evaluated using the metric evolution equations. We note that
  this is the first point we have used the evolution equations for
  $g_{ij}$, and therefore the accuracy with which our 
  spacetime obeys these evolution equations enters can enter into our
  quantities. In other words, if the relationship between $\partial_t
  g$ and $2 \alpha K$ is only obeyed to a given order, we cannot
  expect our quantities which use this relationship to be obeyed at a
  higher order.  By virtue of $\ell^t = 1$,
\begin{equation}
\frac{\partial g_{ij}}{\partial \bar t}\equiv
\ell^\mu \partial_\mu g_{ij} = 
-2\alpha K_{ij} +
D_i \beta_j + D_j \beta_i 
 + \ell^k g_{ij,k}
\end{equation}
where $K_{ij}$ is the extrinsic curvature of the 3 surface. $K_{ij}$
and $g_{ij}$ are both readily available in the numerically constructed
spacetime.

We can find the terms $\partial p^i / \partial \bar t$ by commuting
partial derivatives.  Namely,
\begin{equation}
\frac{\partial p^i}{\partial \bar t} =
\frac{\partial}{\partial \bar t} \frac{\partial x^i}{\partial \bar
\theta} =
\frac{\partial \ell^i}{\partial \bar \theta}.
\end{equation}
The time derivative of $\vec p$ is the spatial derivative of
$\vec\ell$.  We can evaluate the spatial derivative of $\vec\ell$ with
a single time slice finite difference of our surface and surface
quantities, and thus find the required time derivatives. To summarize,
expanding the
time derivative of the horizon metric using the chain rule, and using
the above two techniques, we can find the $\partial \gamma_{ab} /
\partial \bar t$ terms in a single time slice.

Thus, we have a method for finding the four horizon quantities which
describe the kinematics of the horizon surface. This method is
contained entirely in a single 3-slice. We should note that it is also
possible to create the membrane quantities in a direct fashion using
numerical derivatives in time to evaluate the expansion and shear. We
call this the ``time difference'' evaluation of the expansion,
as opposed to the ``single slice'' evaluation.  We find that the single
slice method invariably gives smoother and more accurate data for the
membrane quantities than the time difference method.

An additional difficulty comes in evaluating the horizon equations,
Eq.~(\ref{eq:tidal_force})-(\ref{eq:hajicek}).  Two terms pose a
difficulty there, $D_{\bar t} Z_a$ and ${{\sigma_a}^b}_{||b}$, where
$Z$ is any tensor on the horizon.  Luckily, we only need to evaluate
these terms as a check; we do not use the horizon equations in our
evolution. Thus we can use first order accurate methods to evaluate
these if need be.

We first turn our attention to $D_{\bar t} Z_a$.  First we introduce a
Christoffel symbol for the $(\bar t, \bar\theta, \bar\phi)$
coordinates (e.g., the null horizon 3-surface in co-moving coordinates,
which we will here index  with $(q,r,\ldots)$).
We denote this as ${}^{(3)}{\Gamma^q}_{rs}$. We find
\begin{equation}
D_{\bar t}Z_a = \frac{\partial Z_a}{\partial \bar t} -
        {}^{(3)}{\Gamma^q}_{a \bar t} Z_q.
\label{eq:cov_der_hard}
\end{equation}
The horizon ``3-metric'', $\gamma_{qr}$ is simply given by
 $\gamma_{ab}$ if $q,r \not = \bar t$ and $0$ elsewhere.  Thus
  ${}^{(3)}{\Gamma^q}_{rs}$ can be simply evaluated as
\begin{eqnarray}
{}^{(3)}{\Gamma^q}_{a \bar t} &=& \frac{1}{2}\gamma^{qr}
\left (\gamma_{ra,\bar t} + \gamma_{r\bar t,a} - \gamma_{\bar t a, r}
\right )\nonumber\\
 &=& \frac{1}{2}\gamma^{qb}\gamma_{ab,\bar t}
\end{eqnarray}
Thus we can easily evaluate Eq.~(\ref{eq:cov_der_hard}) as
\begin{equation}
D_{\bar t} Z_a = \frac{\partial Z_a}{\partial \bar t} -
\frac{1}{2}\gamma^{bc} \gamma_{ba,\bar t} Z_a.
\end{equation}
The only term which we cannot calculate in a single slice is $\partial Z_a /
\partial \bar t$, but we can simply calculate that by storing the
quantity $Z_a$ at three time steps and then use a centered
time derivative to evaluate the term at the middle step after all
three steps are taken.

The term ${{\sigma_a}^b}_{||b}$ is evaluated directly, e.g.,
\begin{equation}
 {{\sigma_a}^b}_{||b} = {{\sigma_a}^b}_{,b} + {}^{(2)}{\Gamma^b}_{cb}
{\sigma_a}^c
\end{equation}
and the 4 independent non-zero terms of $ {}^{(2)}{\Gamma^a}_{bc}$ are
evaluated directly from spatial derivatives of the horizon 2-metric.

\subsection{Test and Applications of MP Quantities}
\label{application}
In this section we apply the membrane quantities to a set of testbed
analytical and numerical black hole spacetimes that have been computed
using codes described in Refs. \cite{Abrahams92a,Brandt94b}.  Our aim
here is to probe whether these tools can be used in a
practical way to explore the dynamics of black hole horizons in
numerically generated spacetimes.  We will consider the physics of
these quantities, for a set of interesting spacetimes, in a future
paper. 

\subsubsection{Flat space}

Flat space in Minkowski coordinates,
\begin{equation}
ds^2 = -dt^2 + dr^2 + r^2 d\Omega^2
\end{equation}
allows us to test our expressions for $\expn$ against easily
understandable analytic solutions. Although flat space has no EH, it
does have null surfaces, and our construction carries over to them.

Most notably, we know that for spherical null surfaces in flat space,
the expansion of a sphere of radius $r$ is
\begin{equation}
\expn = \frac{1}{\cal A} \frac{\partial {\cal A}}{\partial t} =
\frac{2}{r}.
\end{equation} 
since in flat space, ${\cal A} = 4 \pi r^2$ and $\partial r / \partial
t = c = 1$. Using this relationship, we can trivially check our
expressions for $\expn$. Additionally, we can form $\partial{\cal A} /
\partial t$ from integrals of the expansion, which carries over into
the dynamical black hole case, where we can compare this integral of
$\expn^H$ with a numerically calculated $\partial A / \partial t$.
Evaluating the expansion in flat space gives the expected
answer.

\subsubsection{Analytic Schwarzschild}

We next turn to the analytic Schwarzschild spacetime described in
standard coordinates,
\begin{equation}
ds^2 = \left ( 1 - \frac {2M}{r}\right) dt^2 + 
\frac{dr^2}{\left( 1 - \frac{2M}{r}\right)} +
r^2 d\Omega^2.
\end{equation}
In this spacetime, the expected results are that the generators and
surface will be attracted backwards in time towards the true horizon
(at $r = 2M$), that $\expn_H$, $\sigma^H_{ab}$ and $\Omega^H_a$
approach zero exponentially as the surface approaches the true
horizon, and that $g_H$ approaches the analytic value of $1/2M$.
Moreover, we can check that relationship between the integral of the
expansion and the area change holds in this spacetime by numerically
differentiating the horizon area, which allows another test of our
expressions. These relationships are obeyed.

In analytic Schwarzschild we can trivially evaluate the above
expressions for $\vec \ell$, $\expn$ and $g_H$ on an arbitrary null
sphere of radius $r$ to find
\begin{eqnarray}
\vec\ell &=& \left(1,1-\frac{2M}{r},0,0\right), \\
\expn &=& \frac{2}{r} \;\ell^r = 
\frac{2}{r} \left (1 - \frac{2M}{r}\right), \\
g_H &=& \frac{2M}{r^2} \; \frac{\ell^t \ell^r}{1 - \frac{2M}{r}} =
\frac{2M}{r^2} .\\
\end{eqnarray} 
Note that $\ell^r$ and $\expn$ vanish on the horizon ($r=2M$) as
expected, and $g_H$ takes the value $1/2M$. We check these
relationships for surfaces away from the horizon and we see that our
surfaces give the analytic results for all null spheres in the spacetime.

Additionally, each of the horizon equations,
Eq.~(\ref{eq:tidal_force})-(\ref{eq:hajicek}) should be obeyed in this
spacetime.  We evaluate only the focusing equation violation, however, since the
tidal force equation contains the electric part of the Weyl tensor,
${\mathcal E}^H_{ab}$, which causes this equation not to be a check on
the membrane quantities alone, and the Hajicek equation is trivially
satisfied with a spherically symmetric $g_H$ and $\Omega^H_a = 0$.

The vanishing of the focusing equation violation allows us a strong check on our
method. Since the focusing equation requires the covariant derivative
of the expansion, $D_t \expn$, we expect the focusing equation to be
obeyed as accurately as $D_t \expn$ is evaluated. Recall, we evaluate
$D_t \expn$ by taking a centered finite difference in time, so we
expect the focusing equation violation in our spacetime to converge
towards zero at ${\cal O}(\Delta t^2)$.  We test this by finding a
surface in the analytic Schwarzschild background first using a Courant
factor $\lambda = 0.2$ and then $\lambda = 0.4$, doubling the time
step. We then measure the focusing equation violation in these two
runs. If the result is converging towards zero, the focusing equation
violation should be four times larger in the $\lambda = 0.4$ case. We
demonstrate this convergence in Fig.~\ref{foceqn_conv} by plotting the
focusing equation violation with $\lambda = 0.2$ as a line, and by
plotting one quarter the focusing equation violation with $\lambda =
0.4$ as diamonds. The demonstration that these two sets of data are
the same indicates that we are converging towards a surface which
satisfies the focusing equations. We note that, as the surface becomes
very close to the actual horizon, the focusing equation is zero at
levels close to machine precision in both simulations, so convergence
can no longer be observed numerically.

%%-FIG-%% foceqn_conv
\vbox{ \begin{figure}
\incpsf{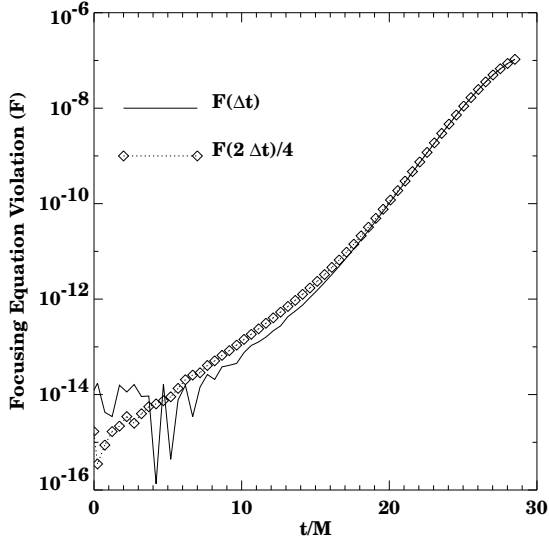}
\caption{We show the violation of the focusing equation, $F$, for a given
  sized time step as a solid line, and one quarter the violation for
  double the time step as diamonds, for a surface integrated in the
  analytic Schwarzschild spacetime. The fact that these data are
  coincident indicates that we are converging towards a null surface
  which satisfies the focusing equation. The surface in question
  starts at $r=2.4M$ at $t=30$. The exponential shrinking of the
  violation is directly due to the exponential approach of
  the expansion towards zero. Note also that as the focusing equation
  approaches machine precision levels (here $10^{-14}$) convergence
  fails, since both quantities are effectively zero.}
\label{foceqn_conv}
\end{figure} }

\subsubsection{Maximally Sliced Schwarzschild}
\label{sec:mpmaxsch}

With the advent of new hyperbolic systems for the Einstein equations
\cite{Bona97a,Arbona98a,Scheel97} and Apparent Horizon Boundary Conditions
\cite{Anninos94e,Daues96a}, long time highly accurate one dimensional
evolutions of a maximally sliced Schwarzschild black hole are quite
readily available, and so we can use these very accurate spacetimes to
test our horizon finding method. In this section, we consider a
maximally sliced black hole evolved with the eigen-method code
described in Ref.~\cite{Arbona98a}, which allows long time evolution with
exceptionally small error.

We first can test the evaluation of the horizon 2-metric,
$\gamma_{ab}$. In the case of no angular generator motion, where the
generators are chosen to be identically on the points on which the
horizon surface $f(t,x^i)$ is evolved, the horizon 2-metric
$\gamma_{ab}$ and the induced surface 2-metric used to evaluate area
and circumferences should be identical. That is, we should get the
same answer evaluating Eq.~(\ref{eqn:sfcarea}) whether we use $\gamma_{ab}$
as defined by Eq.~(\ref{eqn:surfmetric}) or Eq.~(\ref{generator metric}).
Moreover, the vectors $\vec p$ and $\vec q$ should have components $(0,0,1,0)$ and
$(0,0,0,1)$ respectively.  We see both of these features to machine
precision in the maximally sliced Schwarzschild spacetimes.

Spherical symmetry also leads to a vanishing shear; our expression for
the shear vanishes to machine precision. However the expressions for
the expansion is non-trivial, and since we have a very small (but
non-zero) area growth due to numerical error, we can very accurately
measure how well the expansion measures area change.

We choose two trial surfaces for our test, one slightly outside the
horizon and one slightly inside, and integrate them backwards in time.
As expected from Paper I, these surfaces converge towards each other
rapidly, and lock onto the same surface, but have some non-trivial
area change, due to the ``locking on'' process before the surfaces
join the horizon, and due to numerical error afterwards. In Fig.
\ref{maxschwarzadot} we plot $\partial{\cal A}/ \partial t$ calculated
by differentiating the area reported by the code, and also by
integrals over the surface of the expansion. We see that these
quantities agree, strongly indicating that our evaluation of the
expansion is correct. 

%%-FIG-%% maxschwarzadot
\vbox{ \begin{figure}
\incpsf{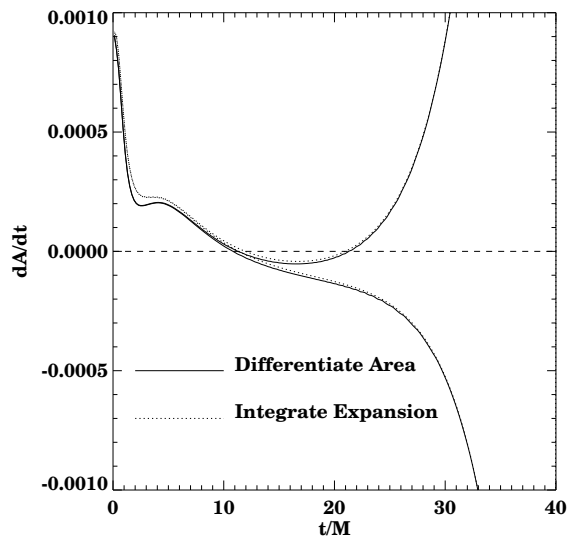}
\caption{We show $\partial{\cal A}/\partial t$ evaluated by taking
  both the numerical derivative of the area calculated by the code,
  and surface integrals of the expansion found from the comoving
  horizon two metric. We use a very accurate maximally sliced
  Schwarzschild spacetime which has a very small, but
  non-zero, numerical error in the spacetime. We integrate
  two surfaces, one originally inside and the other originally outside
  the event horizon. We note the excellent agreement between the two
  measures of $\partial{\cal A}/\partial t$.}
\label{maxschwarzadot}
\end{figure} }

\subsubsection{Small Distortion Non-rotating Black Hole}

We turn to the small distortion Brill wave plus black hole spacetime
considered above.  We first test if our evaluation of the horizon two
metric, $\gamma_{ab}$, gives measures of the horizon geometry which
are consistent with the measures discussed in Sec.\ref{geometry}.
Since the generators will experience angular deflection, integrals to
form areas and circumferences will be over different coordinate
locations when using the comoving and induced two metric. Moreover,
the measure of the geometry using the horizon two metric will be
measured on a non-regular grid in $\theta,\phi$ space (but a regular
grid in $\bar\theta, \bar\phi$ space), and will therefore have an
additional inaccuracy. Nonetheless, we see good agreement. In
Fig.\ref{smalldist_crcomp}, we show the difference in evaluating $C_p$
(not $C_r$) using the comoving and induced two metric in the
Brill-wave plus black hole spacetime. We show the difference for 38
and 76 generators, respectively. Note as the number of generators
increases (therefore reducing numerical error in the integration over
the horizon metric due to generator deflection), the results converge
towards the same solution, or the differences converge towards
zero.

%%-FIG-%% smalldist_crcomp
\vbox{ \begin{figure}
\incpsf{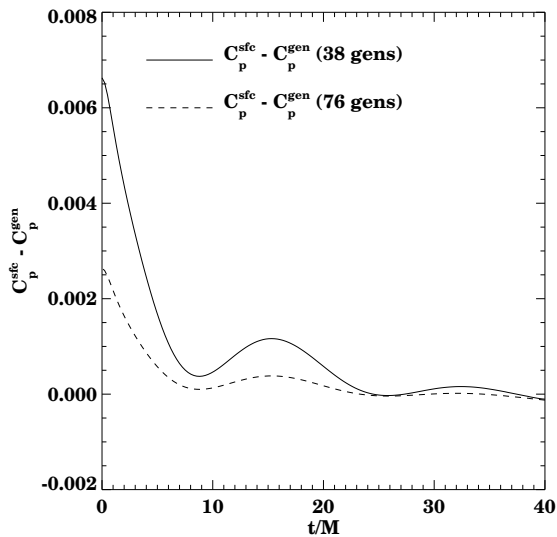}
\caption{We demonstrate that the generator co-moving metric gives
  accurate evaluations of the polar circumference, $C_p$, in the
  spacetime with small amplitude Brill waves initially on the throat.
  We show this by forming $C_{p}^{\mathrm sfc}$ from the induced
  surface metric, and $C_{p}^{\mathrm gen}$ from the co-moving
  generator horizon 2--metric. We taking the difference of the two
  measures with different numbers of generators used to form the
  horizon 2--metric. Clearly, as more generators are used the two methods
  become closer and the differences converge towards zero.}
\label{smalldist_crcomp}
\end{figure} }

We turn next to the expansion on the horizon. For the physical setup
considered here, a gravitational wave incident on a black hole, but
with the wave centered at the throat, we expect the horizon to grow at
$t=0$, and then as time progresses, become static. This should show up
as a positive expansion decreasing towards zero as time progresses.
However, we also know that our spacetime has spurious area growth of
the horizon due to numerical error in the spacetime, as
found in previous studies of the AH. This should appear as a positive,
and increasing, expansion at later times. In Fig.~\ref{exp_evol} we
show the expansion for this spacetime, and see exactly this behavior.
However, a few features of the expansion should be noted. Firstly,
note that at late times, the expansion is not terribly smooth in time.
Secondly, note that, near the axis ($\theta = 0$) the expansion is
somewhat oscillatory. At late time and near the axis the numerically
constructed spacetime is less accurate. We see that our membrane
paradigm quantities as analysis tools are very sensitive detectors of
these errors in the numerically generated spacetime.

%%-FIG-%% exp_evol
\vbox{ \begin{figure}
\incpsf{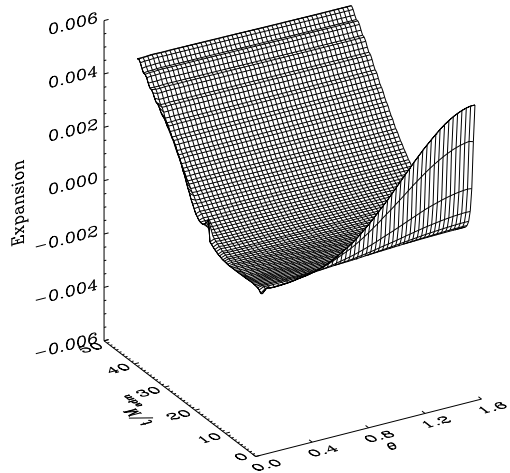}
\caption{We show the evolution of the expansion in time for the
  horizon interacting with a small amplitude Brill wave. Two features
  are of interest here. First, we note the initial expansion is quite
  large but drops quickly, as the horizon swallows the initially
  incident gravitational radiation. This initial growth is
  concentrated near the equator, as the gravitational wave has a
  $\sin^2 \theta$ form. Secondly we note that at later times the
  expansion is growing, as expected from the spurious horizon growth
  due to numerical error, and this growth has no angular dependence.
  We also note a small amount of noise on the horizon near the axis,
  due to spacetime inaccuracies there.}
\label{exp_evol}
\end{figure} }

This detection of error leads us to study how these quantities behave
with changing resolution in the construction of the numerical
spacetime. In Fig.~\ref{exp_reseffect} we take the
same wave parameters used above with resolutions of $200 \times 54$
and $300 \times 80$ to generate two spacetimes. In
Fig.~\ref{exp_reseffect}, we show the area change predicted by
integrating the expansion over the 2-surface. We see that, at $t=0$,
where area change is caused by infalling gravitational radiation and
the spacetime is still quite accurate, both systems give the same
result, but at later times, the expansion due to spurious numerical
error is considerably larger in the lower resolution spacetime, and
the expansion appears to be converging towards zero. In the
high resolution spacetime, the expansion is fairly inaccurate near the
pole, as the system is very susceptible to axis instabilities, but
this noise does not show up in the calculation of area change, as
$\sin\theta$ terms in the integral of the expansion kill this
contribution near the pole.

%%-FIG-%% exp_reseffect
\vbox{ \begin{figure}
\incpsf{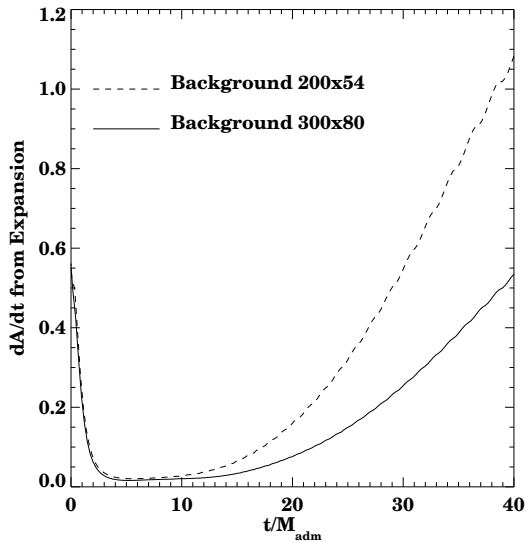}
\caption{We study the behavior of the expansion with a moderate and
  high resolution numerically generated spacetime. At early times
  when area growth is due to accurately modeled gravitational
  phenomena, the expansion should be unchanged by adjusting the
  spacetime. At late times, when area growth is due to
  spurious numerical error, the area growth should is smaller with a
  higher resolution simulation, as the spacetime is converging (at
  roughly second order) towards
  a zero-area-growth solution.}
\label{exp_reseffect}
\end{figure} }

We next turn to the shear. In this spacetime, we expect a non-zero
shear since there is generator motion, but we also expect the shear
$\sigma_{ab}$ to be diagonal, since the spacetime is non-rotating and
axisymmetric. In Figs.~\ref{shear_tt} and \ref{tr_shear} we plot the
evolution of $\sigma_{\bar\theta\bar\theta}$ and the trace of the
shear, ${\sigma_{a}}^a$ in time. We note that the shear is largest
near the equator, and vanishes on the pole, as symmetry arguments
require it must. (There can be no shear at the pole in axisymmetry,
only expansion, since shear at the pole would imply a $\phi$
dependence of the generator motion). We also note that the trace of
the shear vanishes to machine precision in Fig.~\ref{tr_shear}

%%-FIG-%% shear_tt
\vbox{ \begin{figure}
\incpsf{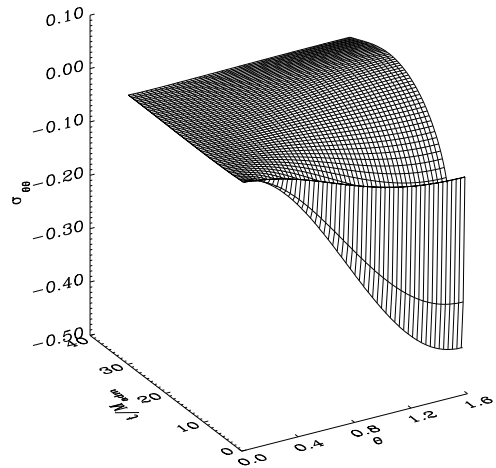}
\caption{We show $\sigma_{\theta\theta}$ on the horizon for the low
  amplitude distortion case considered. We note that there is no shear
  at the poles, and the shear is maximal near the equator.}
\label{shear_tt}
\end{figure} }

%%-FIG-%% tr_shear
\vbox{ \begin{figure}
\incpsf{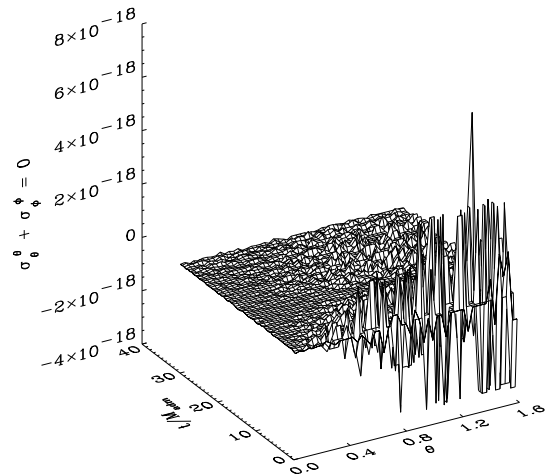}
\caption{We show the trace of the shear, ${\sigma_{\theta}}^\theta 
  + {\sigma_{\phi}}^\phi$ on the horizon for the low amplitude
  distortion case considered. We note that, even though the shear and
  the horizon two metric are order unity, this quantity effectively
  vanishes to machine precision.}
\label{tr_shear}
\end{figure} }

Finally, to test the surface gravity, we turn to the focusing
equation, which is a complicated combination of the surface gravity,
shear, and expansion. If this equation is roughly satisfied in our
spacetime, then we have a strong verification that we are indeed
measuring the membrane quantities appropriately. We test this by
taking the averaged value of the focusing equation violation (or the
LHS - RHS of Eq.~(\ref{eq:focusing})) over the surface. In
Fig.~\ref{foceqn_dyn} we show these averages evolving in time in our
moderate and high resolution spacetimes.  We note that the
focusing equation violation is small, being substantially smaller than the
square of the shear and the expansion.
However it is clear that the evaluation of the focusing equation
violation is also sensitive to the errors in the numerical spacetime
and interpolations.  Noise, which is generated from the discrete and
inaccurate features of the spacetime, is clear in
Fig.~\ref{foceqn_dyn}. However, we also observe that, with more
spacetime resolution, the focusing equation violation converges
towards zero at approximately second order, as expected.

%%-FIG-%% foceqn_dyn
\vbox{ \begin{figure}
\incpsf{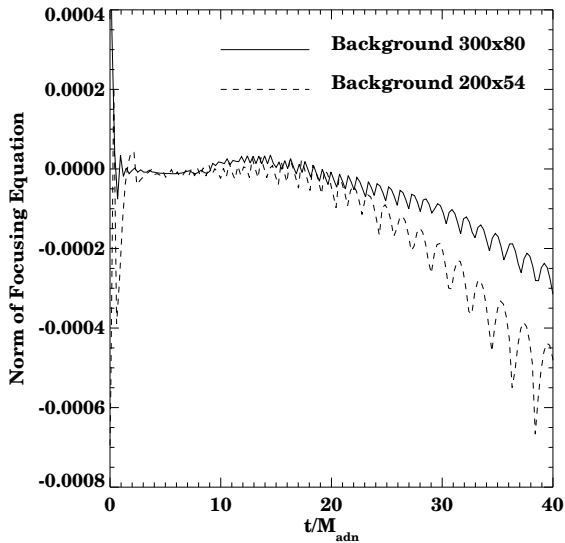}
\caption{We show the norm of the focusing equation over the surface in
  the high resolution and medium resolution spacetimes. We
  note several features. Firstly, this quantity is noisy, but small
  compared to the square of the shear and expansion, both of which
  enter into the equation. Secondly, we note that with an increasingly
  accurate spacetime, the focusing equation converges towards zero at
  approximately second order.}
\label{foceqn_dyn}
\end{figure} }

From the experiments in these two numerical spacetimes we 
conclude that our construction is appropriate for measuring and
generating membrane paradigm type analysis quantities in numerical
spacetimes. These quantities are sensitive detectors of error in
numerical spacetimes, and they allow us to measure detailed properties
of the event horizon and its dynamics. 

\section{Conclusions}

In this paper, we have developed a set of tools with which one can
measure and understand the dynamics of event horizons in numerically
generated spacetimes. We have shown that standard geometric measures
of the horizon are useful tools for understanding horizon dynamics. We
have investigated the behavior of the generators of the horizon in
several spacetimes, including two black hole spacetimes, where
horizons contain caustics, through which generators leave the horizon.
Finally we presented a construction which applies the membrane
paradigm to numerical relativity. We demonstrated that this
construction was effective on analytic spacetimes, and is also
applicable to numerically generated spacetimes. We also note that our
techniques are applicable to any null surface, so could potentially be
useful for studying null surface dynamics in spacetimes without black
holes, or away from black holes. We look forward to more accurate
dynamical black hole spacetimes, so we can use these quantities for
detailed horizon analysis in numerical relativity.

\section{acknowledgements}
We are most grateful to Kip Thorne for many discussions and for
contributing ideas to this work.  We are indebted to Peter Anninos for
his assistance with the two black hole studies and for many useful
conversations.  We are grateful to Steve Brandt for providing data for
evolved rotating black holes that were discussed in this paper, and
for helpful discussions about the Kerr spacetimes. We thank David
Bernstein, David Hobill, and Larry Smarr for providing their
axisymmetric Black Hole + Brill Wave spacetime. We thank Greg Daues
for providing the one-dimensional AHBC spacetime used to create
Fig.~\ref{schwarzarea}. We thank Carles Bona and Joan Stela for
providing the code used in Sec.\ref{sec:mpmaxsch}. We thank Carsten
Gundlach for interesting conversations.

This work used supercomputer facilities at the National Center for
Supercomputing Applications (NCSA), Albert Einstein Institute (AEI),
and the Pittsburgh Supercomputing Center (PSC). This work was
supported by the Albert Einstein Institute, NSF Metacenter Allocation
MCA93025, and NSF grants Nos.  PHY94-04788, PHY94-07882, PHY06-00567,
ASC95-03978 and ASC93-18152.  W.M.S. would like to thank the support
of the Institute of Mathematical Science of the Chinese University of
Hong Kong.

%%%%%%%% Tables %%%%%%%%%%

%%%%%%%% Figures %%%%%%%%%

% Fig 1

% Fig 2

% Fig 3

% Fig 4

% Fig 5

% Fig 6

% Fig 7

% Fig 8

% Fig 9

% Fig 10

% Fig 11

% Fig 12

% Fig 13

% Fig 14

% Fig 15

% Fig 16

% Fig 17

% Fig 18

% Fig 19

% Fig 20

% Fig 21

% Fig 22

% Fig 23

% Fig 24

% Fig 25

% Fig 26

% Fig 27

% Fig 28

\end{document}